\documentclass{ws-ijmpa}
\usepackage[super,compress]{cite}
\usepackage{graphicx}
\usepackage{float}
\begin{document}
\markboth{M. Djouala and N. Mebarki}{Neutral Higgs bosons Phenomenology in the compact 341 model and confrontations with the LHC results}

%

\title{Neutral Higgs bosons phenomenology in the compact 341 model and confrontations with the LHC results
}

\author{M. Djouala\footnote{
Corresponding
author.} 
}

\address{Laboratoire de Physique Math\'{e}matique et Subatomique,\\
	Fr\`{e}res Mentouri University, Constantine1, Algeria	\\
*djoualameriem@gmail.com}

\author{ N. Mebarki
}

\address{Laboratoire de Physique Math\'{e}matique et Subatomique,\\
	Fr\`{e}res Mentouri University, Constantine1, Algeria	\\
	nnmebarki@yahoo.fr}
\maketitle

\begin{history}
\received{Day Month Year}
\revised{Day Month Year}
\end{history}

\begin{abstract}
The phenomenology of the neutral scalar bosons in the compact 341 model is discussed. We show that the predictions of this model are fairly good and compatible with the experimental data by calculating the signal strength of $h_1$  for the channels $b\bar{b}$, $WW^*$, $ZZ^*$, $\tau\bar{\tau}$ and $\gamma\gamma$. Moreover, we use the branching ratios to search for the heavy neutral scalar bosons $h_2$ and $h_3$ where we take into account the theoretical constraints to precise the allowed
ranges for the fundamental parameters of the scalar potential.
\end{abstract}
\ccode{PACS numbers:12.60.-i, 14.80.Cp}

\section{Introduction}
~~~The discovery of a new scalar resonance by the ATLAS and CMS at CERN  with a mass around 125 GeV \cite{Aad,Chatrchyan} in the Run 1  has paved the way for new directions in high energy particle physics.  Analyzing the properties of this particle is compatible with the standard model (SM) predictions, that is a strong evidence that it is the Higgs boson of the Standard Model.
Currently the combined analysis based on the Run 1 (7 and 8 TeV) LHC data shows that the couplings of the SM Higgs boson with the vector bosons are found to be compatible with those expected from the SM within a $\sim$ 10$\%$ uncertainty, whereas the coupling of the SM higgs boson to the third generation fermions is compatible within an uncertainty of $\sim$ 15-20$\%$. Thus, the current status of the Higgs properties still allows to explore new interpretations of the observation coming from new physics of different underlying structures.\\
\indent  The compact 341 model is among many BSM theories \cite{AArhrib, Aulakh,Caetano,Dias:2013kma,Randall,Pisano} which is constructed from the gauge group $SU(3)_C\otimes SU(4)_L \otimes U(1)_X$, it predicts the existence of new particles namely the exotic quarks, scalar and the gauge bosons, moreover, the physical scalar spectrum of this kind of models is composed by three quadruplets $\chi$, $\eta$ and $\rho$ which lead to three neutral CP-even scalars $h_{1,2,3}$ where the lightest neutral scalar boson $h_{1}$ plays the role of the Standard Model Higgs boson, four simply charged $h_{1,2}^{\mp}$ with two doubly charged scalar bosons $h^{\mp\mp}$. The compact 341 model has a very specific arrangement of the fermions into generations: for leptons, one has both right and left handed
helecities arranged in the same multiplet. In order to make the model anomaly free, the second and third quarks families have to belong to the conjugate $4^*$ fundamental representation of the $SU(4)_L$ gauge group, while the first family transforms as a quadruplet in the fundamental representation or vice versa.\\
\indent The motivations to study 341 models are various
as, for example, explanation of family replication \cite{Pisano,Frampton}, electric charge quantization \cite{Sousa98,Sousa99}, strong CP-problem \cite{B.Pal,Dias,G.Dias}, incorporation of inflation\cite{Ferreira} and others \cite{M.Djouala}. Moreover, the 341 models are able to deal with the current Higgs physics which has an important role on the LHC to discover new physics. In this work we  will study the decay of the  neutral scalar bosons in the compact 341 model and confront our results to the experimental data. \\
\indent  The paper is organized as follow: section \ref{sec:Brief review of the model} reviews 
the fundamental features of the compact 341 model by focusing on its particle content and its scalar potential followed by the mass spectrum. In section \ref{sec:Constraints}, we review the theoretical constraints on the
scalar potential parameters which found from the tree-level vacuum stability, perturbative unitarity and from the positivity of the scalar boson masses. All couplings (Feynman rules) needed to calculate the BRs
and signal strengths of the neutral scalar bosons $h_1, h_2, h_3$ are presented in section \ref{sec:Feynman}. In section \ref{sec:Higgs Decays}, 
we calculate the signal strength of $h_{1}$ using the various analytical expressions of the partial
decays width for different channels $\tau^+\tau^-$, $b\bar{b}$, $WW^*$, $ZZ^*$, $\gamma\gamma$ and $Z\gamma$ and the branching ratios of the other heavy scalar bosons $h_{i}$ (i=2,3). We present our numerical results in section \ref{sec:Numerical Analysis}. We draw our conclusions in section \ref{sec:Conclusion}. Finally, the expressions of the theoretical constraints are given in the appendix.
\section{Brief review of the model}
\label{sec:Brief review of the model}
~~~~In general, one of the most important parameters to distinguish different 341 models
are denoted as $\beta$ and $\gamma$ which define the electric charges of new particles through the following electric charge operator \cite{Cogollo:2014tra}:
\begin{equation}
\label{eq:x}
\begin{split}
Q&=\frac{1}{2}\bigg(\lambda_{3}-\frac{1}{\sqrt{3}}\lambda_{8}-\frac{4}{\sqrt{6}}\lambda_{15}\bigg)+X\,.
\end{split}
\end{equation}
Where
\begin{equation}
\lambda_{3}=\text{diag}(1,-1,0,0)\,,~ \lambda_{8} =
\frac{1}{\sqrt{3}}\text{diag}(1,1,-2,0) \,,~
\lambda_{15} = \frac{1}{\sqrt{6}}\text{diag}(1,1,1,-3) \,.
\end{equation}
\label{sec:intro}
The particle content of this model is represented in the following table
\cite{Dias:2013kma}:
\begin{table}[H]
	\tbl{\label{tab:mld} The complete anomaly free particle content of the compact 341 model where the symbol $\sim$ refers to the quantum
		numbers of the $SU(3)_{C}$, $SU(4)_{L}$, $U(1)_{X}$ respectively
		 and F refers to flavors}
	{\begin{tabular}{@{}ccccccccccccccc@{}} \toprule
			Name & 341 rep& 331 rep & components&F \\ \colrule
			$\psi_{\alpha L}$\hphantom{00}&(1,4,$0$)\hphantom{00}&(1,3,$\frac{-1}{3}$)+(1,1,1)\hphantom{00}&($\nu_{\alpha},l_{\alpha},\nu_{\alpha}^{c},l_{\alpha}^{c}$)&3\\
			$Q_{iL}(i=2,3)$&(3,$\overline{4}$,$\frac{-1}{3})$&(3,$\bar{3}$,0)+(3,1,$\frac{-4}{3}$)&$(d_{i},u_{i},D_{i},J_{i})$&2\\
			$Q_{1L}$&(3,4,$\frac{2}{3})$&(3,3,$\frac{1}{3}$)+(3,1,$\frac{5}{3}$)&$(u_{1},d_{1},U_{1},J_{1})$&1\\
			$u_{jR}$(j=1,i)&($\overline{3}$,1,$\frac{2}{3}$)&($\overline{3}$,1,$\frac{2}{3}$)&$u_{jR}$&4\\
			$d_{jR}(j=1,i)$&($\overline{3}$,1,$\frac{-1}{3}$)&($\overline{3}$,1,$\frac{-1}{3}$)&$d_{jR}$&5\\
			$J_{1R}$&($\overline{3}$,1,$\frac{5}{3}$)&($\overline{3}$,1,$\frac{5}{3}$)&$J_{1R}$&1\\
			$J_{iR}(i=2,3)$&($\overline{3}$,1,$\frac{-4}{3}$)&($\overline{3}$,1,$\frac{-4}{3}$)&$J_{iR}$&2\\
			$\chi$&(1,4,-1)&(1,3,
			$\frac{-4}{3}$)+(1,1,0)&($\chi_{1}^{-}$,$\chi^{--}$,$\chi_{2}^{-}$,$\chi^{0})$&1\\
			$\eta$&(1,4,0)&(1,3,$\frac{-1}{3}$)+(1,1,1)&($\eta_{1}^{0}$,$\eta_{1}^{-}$,$\eta_{2}^{0}$,$\eta_{2}^{+})$&1\\
			$\rho$&(1,4,1)&(1,3,$\frac{2}{3}$)+(1,1,2)&($\rho_{1}^{+}$,$\rho^{0}$,$\rho_{2}^{+}$,$\rho^{++})$&1\\ \botrule
		\end{tabular} \label{ta1}}
\end{table}
To generate masses for all particles, three quadruplets scalar fields $SU(4)_L$  are introduced as we indicated in table \ref{tab:mld} where:
\begin{eqnarray}
\chi^{0}&=&\frac{1}{\sqrt{2}}(\upsilon_{\chi}+R_{\chi}+iI_{\chi}),\\
\eta^{0}&=&\frac{1}{\sqrt{2}}(\upsilon_{\eta}+R_{\eta}+iI_{\eta}),\\
\rho^{0}&=&\frac{1}{\sqrt{2}}(\upsilon_{\rho}+R_{\rho}+iI_{\rho}).
\end{eqnarray}
with
\begin{equation}
R_{\chi}=h_{1},~~
R_{\eta}=\alpha h_{2}+\beta h_{3},~~
R_{\rho}=\gamma h_{2}+\sigma h_{3},~~
\end{equation}
where $\alpha$, $\beta$, $\gamma$ and $\sigma$ are real parameters:
\begin{eqnarray}
\alpha&=&\frac{-\sqrt{X^{2}+(Y-\sqrt{X^{2}+Y^{2}})^{2}}}{\sqrt{4(X^{2}+Y^{2})}},~~
\beta=\frac{\sqrt{X^{2}+(Y+\sqrt{X^{2}+Y^{2}})^{2}}}{\sqrt{4(X^{2}+Y^{2})}}\\
\gamma&=&\frac{(Y+\sqrt{X^{2}+Y^{2}})(\sqrt{X^{2}+(Y-\sqrt{X^{2}+Y^{2}})^{2}})}{X\sqrt{4(X^{2}+Y^{2})}}\\
\sigma&=&\frac{-(Y-\sqrt{X^{2}+Y^{2}})(\sqrt{X^{2}+(Y+\sqrt{X^{2}+Y^{2}})^{2}})}{X\sqrt{4(X^{2}+Y^{2})}}
\end{eqnarray}
with
\begin{equation}
X=\lambda_{5},~~~~ Y=\lambda_{1}-\lambda_{3},
\end{equation}
and $I_{\chi}$, $I_{\eta}$ and $I_{\rho}$ represent the Goldostone bosons.\\
\indent The most
general scalar potential in the compact 341 model \footnote{This scalar potential is invariant under an additional discrete symmetry $Z_{3}$ \cite{Dias:2013kma}} is given by\cite{Dias:2013kma}: 
\begin{eqnarray}\label{eq:scalarV}
V(\eta,\rho,\chi)&=&\mu_{\eta}^{2}\eta^{\dag}\eta+\mu_{\rho}^{2}\rho^{\dag}\rho+\mu_{\chi}^{2}\chi^{\dag}\chi+\lambda_{1}(\eta^{\dag}\eta)^{2}+\lambda_{2}(\rho^{\dag}\rho)^{2}+\lambda_{3}(\chi^{\dag}\chi)^{2}\nonumber \\
&+&\lambda_{4}(\eta^{\dag}\eta)(\rho^{\dag}\rho)+\lambda_{5}(\eta^{\dag}\eta)(\chi^{\dag}\chi)+\lambda_{6}(\rho^{\dag}\rho)(\chi^{\dag}\chi)+\lambda_{7}(\rho^{\dag}\eta)(\eta^{\dag}\rho)
\nonumber
\\&+&\lambda_{8}(\chi^{\dag}\eta)(\eta^{\dag}\chi)+\lambda_{9}(\rho^{\dag}\chi)(\chi^{\dag}\rho),
\end{eqnarray}
where $\mu^{2}_{\mu\rho\chi}$ are the mass dimension parameters and
$\lambda_{S}$ are dimensionless real coupling constants. \\
\indent In the compact 341 model, the extension of the electroweak  group leads to the existence of nine scalar bosons, namely three neutral scalars $h_{i}$ (i=1,2,3) where the lightest one $h_{1}$ is assumed to be the observed scalar of mass $\sim$ 125 GeV (Higgs boson) at the LHC, while, the other scalars $h_{i}$ (i=2,3) should be heavier, four simply charged $h_{1}^{\mp}$, $h^{\mp}_{2}$ scalars and two doubly charged $h^{\mp\mp}$ ones. After SSB the scalar bosons acquire their masses \cite{Dias:2013kma}:
\begin{eqnarray}
M^{2}_{h_{1}}&\approx&
\lambda_{2}\upsilon_{\rho}^{2}+\frac{\lambda_{3}\lambda^{2}_{4}+\lambda_{6}(\lambda_{1}\lambda_{6}-\lambda_{4}\lambda_{5})}{\lambda^{2}_{5}-4\lambda_{1}\lambda_{3}}\upsilon_{\rho}^{2},~~
M_{h_{2}}^{2}\approx
c_{1}\upsilon_{\chi}^{2}+c_{2}\upsilon_{\rho}^{2}\approx c_{1}\upsilon_{\chi}^{2},\\
M_{h_{3}}^{2}&\approx& c_{3}\upsilon_{\chi}^{2}+c_{4}\upsilon_{\rho}^{2}\approx c_{3}\upsilon_{\chi}^{2},~~
M^{2}_{h^{\mp}_{1}}=\frac{\lambda_{7}}{2}(\upsilon^{2}_{\eta}+\upsilon^{2}_{\rho}),~~
M^{2}_{h^{\mp}_{2}}=\frac{\lambda_{8}}{2}(\upsilon^{2}_{\eta}+\upsilon^{2}_{\chi}),\\
M^{2}_{h^{\mp\mp}}&=&\frac{\lambda_{9}}{2}(\upsilon^{2}_{\rho}+\upsilon^{2}_{\chi}),
\end{eqnarray}
where
\begin{eqnarray}
c_{1}&=&\frac{1}{2}\bigg(\lambda_{1}+\lambda_{3}-\sqrt{(\lambda_{1}-\lambda_{3})^{2}+\lambda_{5}^{2}}
\bigg)\upsilon_{\chi},\\
c_{3}&=&\frac{1}{2}\bigg(\lambda_{1}+\lambda_{3}+\sqrt{(\lambda_{1}-\lambda_{3})^{2}+\lambda_{5}^{2}}
\bigg)\upsilon_{\chi}.
\end{eqnarray}
\indent This model contains 15 electroweak gauge bosons, four of them are identified as the SM gauge bosons and the remaining ones are heavy where some of them are  electrically charged and the others are neutral: $K^{0}$,$K^{\prime0}$,$K^{\mp}_{1}$,$X^{\mp}$,$V^{\mp\mp}$ and $Y^{\mp}$
,$Z^{\prime}$ and $Z^{\prime\prime}$, their masses are found from the following Lagrangian:
\begin{equation}\label{eq:scalarV2}
\mathcal{L}=(D_{\mu}\eta)^{\dagger}(D^{\mu}\eta)+(D_{\mu}\rho)^{\dagger}(D^{\mu}\rho)+(D_{\mu}\chi)^{\dagger}(D^{\mu}\chi),
\end{equation}
where
\begin{equation}
D_{\mu}=\partial_{\mu}+igW_{\mu}^{a}\frac{\lambda_{a}}{2}+iXg_{X}W_{\mu}^{X}, \qquad\text{with}\qquad a=1....15. 
\end{equation}
And by using the following combination:
\begin{eqnarray}
W^{\mp}=\frac{(W_{\mu}^{1}\pm
	iW_{\mu}^{2})}{\sqrt{2}},~~
K^{0},K^{\prime0}=\frac{(W_{\mu}^{4}\mp iW_{\mu}^{5})}{\sqrt{2}},~~
K^{\mp}_{1}=\frac{(W_{\mu}^{6}\mp iW_{\mu}^{7})}{\sqrt{2}},\\
X^{\mp}=\frac{(W_{\mu}^{9}\mp iW_{\mu}^{10})}{\sqrt{2}},~~
V^{\mp\mp}=\frac{(W_{\mu}^{11}\mp iW_{\mu}^{12})}{\sqrt{2}},~~
Y^{\mp}=\frac{(W_{\mu}^{13}\mp iW_{\mu}^{14})}{\sqrt{2}}.
\end{eqnarray}
We get:
\begin{eqnarray}
M_{W{\mp}}^{2}&=&\frac{g^{2}}{4}\upsilon^{2}_{\rho},~~
M_{K^{\prime0},K^{0}}^{2}=\frac{g^{2}}{4}\upsilon^{2}_{\eta},~~
M_{K_{1}^{\mp}}^{2}=\frac{g^{2}}{4}\upsilon^{2}_{\eta},~~
M_{X^{\mp}}^{2}=\frac{g^{2}}{4}\upsilon^{2}_{\chi},\\
M_{V^{\mp\mp}}^{2}&=&\frac{g^{2}}{4}\upsilon^{2}_{\chi},~~
M_{Z}^{2}=\frac{g^{2}\upsilon_{\rho}^{2}}{4c_{W}^{2}},~~
M_{Y^{\mp}}^{2}=\frac{g^{2}}{4}(\upsilon^{2}_{\eta}+\upsilon^{2}_{\chi}),\\
M_{Z^{\prime}}^{2}&=&\frac{g^{2}c_{W}^{2}\upsilon_{\eta}^{2}}{h_{W}},~~ M_{Z^{\prime\prime}}^{2}=\frac{g^{2}\upsilon_{\eta}^{2}\bigg((1-4s_{W}^{2})^{2}+h_{W}^{2}\bigg)}{8h_{W}(1-4s_{W}^{2})}.
\end{eqnarray}
\section{Constraints}
\label{sec:Constraints}
The theoretical constraints on the scalar parameters in the compact 341 model are \cite{Djouala:2020wjj}: 
\begin{itemize}
	\item to ensure the perturbative unitarity, in order to keep the scalar potential bounded from below in the large fields limit for all possible directions in the field space (the vacuum stability) constraints and to provide the vacuum configuration $\langle \rho \rangle_{0}$, $\langle \chi \rangle_{0}$ and $\langle \eta \rangle_{0}$ to be a minimum of the scalar potential one needs to maintain the conditions which are given in the appendix A \cite{Djouala:2020wjj}.
	\item To maintain the perturbativity of the scalar potential, all the scalar quartic couplings $\lambda_{ 1...9}$ have to satisfy:
	\begin{align}
	|\lambda_{i}|\leq4\pi
	\end{align}
	\item The positivity of the scalar bosons masses requires the following conditions:
	\begin{eqnarray}
	\lambda_{1}&+&\lambda_{3}-\sqrt{(\lambda_{1}-\lambda_{3})^{2}+\lambda_{5}^{2}}>0,~~
	\lambda_{1}+\lambda_{3}+\sqrt{(\lambda_{1}-\lambda_{3})^{2}+\lambda_{5}^{2}}>0,\nonumber\\
	\lambda_{2}&+&\frac{\lambda_{3}\lambda^{2}_{4}+\lambda_{6}(\lambda_{1}\lambda_{6}-\lambda_{4}\lambda_{5})}{\lambda^{2}_{5}-4\lambda_{1}\lambda_{3}}>0,~~
	\lambda_{7}>0,~~\lambda_{8}>0,~~ \lambda_{9}>0.
	\end{eqnarray}
	\item A new constraint over the parameters of the scalar potential is imposing by identifying the lightest scalar $h_{1}$ as the Standard Model Higgs boson, then we get:
	\begin{equation}
	\lambda_{2}+\frac{\lambda_{3}\lambda^{2}_{4}+\lambda_{6}(\lambda_{1}\lambda_{6}-\lambda_{4}\lambda_{5})}{\lambda^{2}_{5}-4\lambda_{1}\lambda_{3}}=\frac{m^{2}_{h_{1}}}{\upsilon^{2}_{\rho}},
	\end{equation}
	where we take $\upsilon^{2}_{\rho}$=246 GeV.
	\item Moreover, the existence of a Landau pole in our model $\Lambda$ at a scale $\mu$ \cite{Dias:2013kma},
 all the VEVs and the particle masses $m_i$ are constrained by:
	\begin{equation}
	m_i~~\text{ and VEVs}\leq\Lambda
	\end{equation}
\end{itemize}
\section{Feynman rules}
\label{sec:Feynman}
~~From the scalar potential \eqref{eq:scalarV}, we can derive all Higgs couplings $h_{j}ss$ where $h_{j}$ (j=1,2,3) and s=$h_{1}^{\mp},h_{2}^{\mp},h^{\mp\mp}$ and based on the following Yukawa Lagrangian:
\begin{eqnarray}
\mathcal{L}_{Y}&=&\lambda_{11}^{J}\bar{Q}_{1L}\chi
J_{1R}+\lambda_{ij}^{J}\overline{Q}_{iL}\chi^{*}J_{jR}+\lambda_{1a}^{d}\overline{Q}_{1L}\rho
d_{aR} +\lambda_{ia}^{u}\overline{Q}_{iL}\rho^{*} u_{aR}\nonumber
\\&+&\lambda_{11}^{U}\overline{Q}_{1L}\eta
U_{1R}+\lambda_{ij}^{D}\overline{Q}_{iL}\eta^{*}D_{jR}
+\frac{\lambda_{1a}^{u}}{\Lambda^{2}}\epsilon_{mnop}\bigg(\overline{Q}_{1Lm}\rho^{*}_{n}\chi^{*}_{o}\eta^{*}_{p}\bigg)u_{aR}\nonumber \\
&+&\frac{\lambda_{ia}^{d}}{\Lambda^{2}}\epsilon_{mnop}\bigg(\overline{Q}_{iLm}\rho_{m}\chi_{o}\eta_{p}\bigg)d_{aR}+h.c,
\end{eqnarray}
the couplings of the Higgs bosons $h_{j}$
with fermions $h_{j}\bar{f}f$ can be determined. Furthermore, the couplings of Higgs bosons $h_{j}$ and gauge bosons are contained in the covariant kinetic terms of the Higgs bosons \eqref{eq:scalarV2}.
Whereas,
The triple couplings of three gauge bosons arise from the covariant kinetic Lagrangian of the non-Abelian gauge bosons (Yang Mills Lagrangian):
\begin{equation}
\mathcal{L}=-\frac{1}{4}W^{a}_{\mu\nu}W^{a\mu\nu},
\end{equation}
with
\begin{equation}
W^{a\mu\nu}=\partial^{\mu}W^{a\nu}-\partial^{\nu}W^{a\mu}+gf^{abc}W^{b\mu}W^{c\nu},
\end{equation}
where a,b,c go from 1 to 15 and $f^{abc}$ are the structure constants of the SU(4) group. The non vanishing Feynman rules are listed in the following tables \footnote{In those tables, we mention only couplings that contribute in our discussion.}:
\begin{table}[H]
	\tbl{\label{tab:i8}Higgs $h_{1}$ interactions.}
	{\begin{tabular}{@{}ccccccccccccccc@{}} \toprule
			Interactions & Couplings\\  \colrule 
			$\overline{u}_{i}u_{i}h_{1}$ & $\frac{M^{u_{i}}}{\upsilon_{\rho}}$~\text{with}~$u_{i}\equiv u,c,t$\\
			$\overline{d}_{i}d_{i}h_{1}$ &$\frac{M^{d_{i}}}{\upsilon_{\rho}}~\text{with}~d_{i}\equiv d,s,b$  \\
			$\overline{\ell}\ell h_{1}$&$\frac{m_{\ell}}{\upsilon_{\rho}}$ \\
			$W^{+}W^{-}h_{1}$ &$\frac{g^{2}}{2}\upsilon_{\rho}$ \\
			$ K^{-}_{1}K^{+}_{1}h_{1}$& $\frac{g^{2}}{2}\upsilon_{\rho}$  \\
			$ V^{++}V^{--}h_{1}$ &$\frac{g^{2}}{2}\upsilon_{\rho}$  \\
			$h^{++}h^{--}h_{1}$&$2\lambda_{2}\upsilon_{\rho}$ \\
			$h^{+}_{1}h^{-}_{1}h_{1}$&$2\lambda_{2}\upsilon_{\rho}$ \\
			$ h^{+}_{2}h^{-}_{2}h_{1}$&$\frac{\upsilon_{\rho}}{2}(\lambda_{6}+\lambda_{4})$\\
			$ZZh_{1}$&$\frac{g^{2}\upsilon_{\rho}}{4}\bigg((C^{21})^{2}-\frac{2}{\sqrt{3}}C^{21}C^{22}-\frac{3}{\sqrt{6}}C^{21}C^{23}+\frac{(C^{22})^{2}}{3}+\frac{2}{\sqrt{18}}C^{22}C^{23}$+\\
			&$\frac{(C^{23})^{2}}{6}
			+\frac{4t^{2}}{4}(C^{24})^{2}-$$4tC^{24}C^{21}+4tC^{24}C^{22}
			+\frac{4t}{\sqrt{6}}C^{24}C^{23}\bigg)$\\
			\botrule
	\end{tabular}}
\end{table}
\begin{table}[H]
	\tbl{\label{tab:i9}Higgs $h_{2}$ interactions.}
	{\begin{tabular}{@{}ccccccccccccccc@{}} \toprule
			Interactions & Couplings\\ \colrule
			$\overline{q}qh_{2}$&  $\frac{M^{q}}{\upsilon_{\chi}}(\gamma+\frac{\upsilon_{\chi}}{\upsilon_{\eta}}\alpha)$ \text{with} $q\equiv u,s,b$\\
			$\overline{J}Jh_{2}$ & $\frac{M^{J}}{\upsilon_{\chi}}\gamma$ \\
			$\overline{U}Uh_{2}$&$\frac{M^{U}}{\upsilon_{\eta}}\alpha$ \\
			$\overline{\ell}\ell h_{2}$&$\frac{m_{\ell}\gamma}{\upsilon_{\chi}}$ \\
			$X^{+}X^{-}h_{2}$ &$\frac{g^{2}}{2}\upsilon_{\chi}\gamma$ \\
			$ V^{++}V^{--}h_{2}$ & $\frac{g^{2}}{2}\upsilon_{\chi}\gamma$  \\
			$ K^{\prime0}K^{\prime0}h_{2}$&$\frac{g^{2}}{2}\upsilon_{\eta}\alpha$ \\
			$ K^{-}_{1}K^{+}_{1}h_{2}$&$\frac{g^{2}}{2}\upsilon_{\eta}\alpha$  \\
			$ Y^{-}Y^{+}h_{2}$& $\frac{g^{2}}{2}(\upsilon_{\chi}\gamma+\upsilon_{\eta}\alpha)$  \\
			$h^{++}h^{--}h_{2}$&$\lambda_{4}\upsilon_{\eta}\alpha+(\lambda_{6}+\lambda_{9})\upsilon_{\chi}\gamma$ \\
			$h^{+}_{1}h^{-}_{1}h_{2}$& $(\lambda_{4}+\lambda_{7})\upsilon_{\eta}\alpha+\lambda_{6}\upsilon_{\chi}\gamma$ \\
			$h^{+}_{2}h^{-}_{2}h_{2}$&$\frac{\upsilon_{\chi}\gamma}{2}(2\lambda_{3}+\lambda_{5}+\lambda_{8})+\frac{\upsilon_{\chi}\alpha\lambda_{8}}{2}+\frac{\upsilon_{\eta}\alpha}{2}(
			\lambda_{5}+\lambda_{8}+2\lambda_{1})+\frac{\gamma\upsilon_{\eta}\lambda_{8}}{2}$\\
			$ZZh_{2}$&$\frac{g^{2}\gamma}{4}\upsilon_{\chi}\bigg(\frac{3}{2}(C^{23})^{2}+4t^2(C^{24})^{2}+\frac{12t}{\sqrt{6}}C^{23}C^{24}\bigg) +\frac{g^{2}\alpha}{4}
			\upsilon_{\eta}\bigg(\frac{4}{3}(C^{22})^{2}
			+\frac{(C^{23})^{2}}{6}-\frac{4}{\sqrt{18}}C^{22}C^{23}\bigg)$\\
			$Z^{\prime}Z^{\prime}h_{2}$&$\frac{g^{2}\gamma}{4}\upsilon_{\chi}\bigg(\frac{3}{2}(C^{33})^{2}+4t^2(C^{34})^{2}+\frac{12t}{\sqrt{6}}C^{33}C^{34}\bigg) +\frac{g^{2}\alpha}{4}
			\upsilon_{\eta}\bigg(\frac{4}{3}(C^{32})^{2}
			+\frac{(C^{33})^{2}}{6}-\frac{4}{\sqrt{18}}C^{32}C^{33}\bigg)$\\
			$h_{1}h_{1}h_{2}$&$\frac{\lambda_{6}}{2}\upsilon_{\chi}\gamma+\frac{\lambda_{4}}{2}\upsilon_{\eta}\alpha$\\
			\botrule
	\end{tabular}}
\end{table}

\begin{table}[H]
	\tbl{\label{tab:i10}Higgs $h_{3}$ interactions.}
	{\begin{tabular}{@{}ccccccccccccccc@{}} \toprule
			Interactions & Couplings\\  \colrule
			$\overline{q}qh_{3}$&$\frac{M^{q}}{\upsilon_{\chi}}(\sigma+\frac{\upsilon_{\chi}}{\upsilon_{\eta}}\beta)$~\text{with} $q\equiv u,s,b$\\
			$\overline{J}Jh_{3}$ & $\frac{M^{J}}{\upsilon_{\chi}}\sigma$ \\
			$\overline{U}Uh_{3}$&$\frac{M^{U}}{\upsilon_{\eta}}\beta$ \\
			$\overline{\ell}\ell h_{3}$&$\frac{m_{\ell}}{\upsilon_{\chi}}\sigma$ \\
			$X^{+}X^{-}h_{3}$ &$\frac{g^{2}}{2}\upsilon_{\chi}\sigma$ \\
			$ V^{++}V^{--}h_{3}$ & $\frac{g^{2}}{2}\upsilon_{\chi}\sigma$  \\
			$ K^{\prime0}K^{0\prime}h_{3}$& $\frac{g^{2}}{2}\upsilon_{\eta}\beta$  \\
			$ K^{-}K^{+}h_{3}$& $\frac{g^{2}}{2}\upsilon_{\eta}\beta$  \\
			$ Y^{-}Y^{+}h_{3}$& $\frac{g^{2}}{2}(\upsilon_{\chi}\sigma+\upsilon_{\eta}\beta)$  \\
			$h^{++}h^{--}h_{3}$&$\lambda_{4}\upsilon_{\eta}\beta+\upsilon_{\chi}\sigma(\lambda_{6}+\lambda_{9})$ \\
			$h^{+}_{1}h^{-}_{1}h_{3}$&
			$\upsilon_{\eta}\beta(\lambda_{4}+\lambda_{7})+\lambda_{6}\upsilon_{\chi}\sigma$
			\\$ h^{+}_{2}h^{-}_{2}h_{3}$&
			$\frac{\upsilon_{\chi}\sigma}{2}(2\lambda_{3}+\lambda_{8}+\lambda_{5})+\frac{\upsilon_{\chi}\lambda_{8}}{2}\beta+\frac{\upsilon_{\eta}\beta}{2}(\lambda_{5}+\lambda_{8}+2\lambda_{1})+\frac{\upsilon_{\eta}\sigma\lambda_{8}}{2}$\\
			$ZZh_{3}$&$\frac{g^{2}\upsilon_{\chi}\sigma}{4}\bigg(\frac{3}{2}(C^{23})^{2}+4t^2(C^{24})^{2}+
			\frac{12t}{\sqrt{6}}C^{23}C^{24}\bigg)+\frac{g^{2}\upsilon_{\eta}\beta}{4}\bigg(\frac{4}{3}(C^{22})^{2}+\frac{(C^{23})^{2}}{6}-\frac{4}{\sqrt{18}}C^{22}C^{23}\bigg)$\\
			$Z^{\prime}Z^{\prime}h_{3}$&$\frac{g^{2}\upsilon_{\chi}\sigma}{4}\bigg(\frac{3}{2}(C^{33})^{2}+4t^2(C^{34})^{2}+
			\frac{12t}{\sqrt{6}}C^{33}C^{34}\bigg)+\frac{g^{2}\upsilon_{\eta}\beta}{4}\bigg(\frac{4}{3}(C^{32})^{2}+\frac{(C^{33})^{2}}{6}-\frac{4}{\sqrt{18}}C^{32}C^{33}\bigg)$\\
			$h_{1}h_{1}h_{3}$&$\upsilon_{\chi}\frac{\lambda_{6}}{2}\sigma+\frac{\lambda_{4}}{2}\upsilon_{\eta}\beta$\\
			$h_{2}h_{2}h_{3}$&$\frac{\lambda_{5}\upsilon_{\chi}}{2}(\alpha^{2}\sigma+2\alpha\beta\gamma)+\frac{\lambda_{5}\upsilon_{\eta}}{2}(\beta\gamma^{2}+2\alpha\gamma\sigma)$\\
			$h_{2}h_{1}h_{3}$&$\lambda_{4}\upsilon_{\rho}\alpha\beta+\lambda_{6}\upsilon_{\rho}\gamma\sigma$\\
			\botrule
	\end{tabular}}
\end{table}
\begin{table}[H]
	\tbl{\label{tab:i11} Couplings $\frac{g\gamma^{\mu}}{4}(g_{V}-\gamma_{5}g_{A})$ of
		the Z$q\overline{q}$ vertex.}
	{\begin{tabular}{@{}ccccccccccccccc@{}} \toprule
			Interactions & $g_{V}$ & $g_{A}$\\ \colrule
			$\overline{u}uZ$&$\frac{-5}{3}S_{W}T_{W}+C_{W}$&$\frac{1}{C_{W}}$\\
			$\overline{d}dZ$&$\frac{1}{3}S_{W}T_{W}-C_{W}$&$\frac{-1}{C_{W}}$\\
			$\overline{s}sZ$&$C_{W}-\frac{5}{3}S_{W}T_{W}$&$C_{W}-3S_{W}T_{W}$\\
			$\overline{b}bZ$&$C_{W}-\frac{5}{3}S_{W}T_{W}$&$C_{W}-3S_{W}T_{W}$\\
			$\overline{c}cZ$&$-C_{W}+\frac{1}{3}S_{W}T_{W}$&$-\frac{1}{C_{W}}$\\
			$\overline{t}tZ$&$-C_{W}+\frac{1}{3}S_{W}T_{W}$&$-\frac{1}{C_{W}}$\\
			$\overline{U_1}U_1Z$&$\frac{-8}{3}S_{W}T_{W}$&0\\
			$\overline{J_1}J_1Z$&$\frac{-20}{3}S_{W}T_{W}$&0\\
			$\overline{D_2}D_2Z$&$\frac{4}{3}S_{W}T_{W}$&0\\
			$\overline{D_3}D_3Z$&$\frac{4}{3}S_{W}T_{W}$&0\\
			$\overline{J_2}J_2Z$&$\frac{16}{3}S_{W}T_{W}$&0\\
			$\overline{J_3}J_3Z$&$\frac{16}{3}S_{W}T_{W}$&0\\
			\botrule
	\end{tabular}}
\end{table}

\begin{table}[H]
	\tbl{\label{tab:i12} Couplings ZVV, where V is the gauge bosons $X^{\mp}$, $Y^{\mp}$, $V^{\mp\mp}$, $K^{\mp}_1$, $K'^{0}$ and $W^{\mp}$, while, $\sum_{\alpha\beta\mu}(p,k,q)=g_{\alpha\beta}(p-k)_{\mu}+g_{\beta\mu}(k-q)_{\alpha}+g_{\mu\alpha}(q-p)_{\beta}$.}
	{\begin{tabular}{@{}ccccccccccccccc@{}} \toprule
			Interactions & Couplings\\ \colrule
			$W^{+}_{\beta}W^{-}_{\alpha}Z_{\mu}$&$igC^{21}\sum_{\alpha\beta\mu}(p,k,q)$\\
			$K^{\prime0}_{\beta}K^{\prime0}_{\alpha}Z_{\mu}$&$\frac{-ig}{2}(C^{21}+\sqrt{3}C^{22})\sum_{\alpha\beta\mu}(p,k,q)$\\
			$X^{+}_{\beta}X^{-}_{\alpha}Z_{\mu}$&$\frac{-ig}{2\sqrt{3}}(\sqrt{3}C^{21}+C^{22}+C^{23})\sum_{\alpha\beta\mu}(p,k,q)$\\
			$K^{+}_{1\beta}K^{-}_{1\alpha}Z_{\mu}$&$\frac{-ig}{2}(\sqrt{3}C^{22}-C^{21})\sum_{\alpha\beta\mu}(p,k,q)$\\
			$V^{++}_{\beta}V^{--}_{\alpha}Z_{\mu}$&$\frac{-ig}{2\sqrt{3}}(-\sqrt{3}C^{21}+C^{22}+C^{23})\sum_{\alpha\beta\mu}(p,k,q)$\\
			$Y^{+}_{\beta}Y^{-}_{\alpha}Z_{\mu}$&$\frac{-ig}{2\sqrt{3}}(-C^{22}+C^{23})\sum_{\alpha\beta\mu}(p,k,q)$\\		
			\botrule
	\end{tabular}}
\end{table}
Where 
\begin{eqnarray}
C^{21}&=&C_{W},\qquad
C^{22}=\frac{S_{W}T_{W}}{\sqrt{3}},\qquad
C^{23}=\frac{4S_{W}T_{W}}{\sqrt{6}},\qquad
C^{24}=\frac{-S_{W}T_{W}}{t},\\
C^{32}&=&\sqrt{\dfrac{3-T^{2}_{W}}{3}},\qquad C^{33}=\frac{-4}{\sqrt{6}}\frac{T^{2}_{W}}{\sqrt{3-T^{2}_{W}}},\qquad C^{34}=\frac{T^{2}_{W}}{t\sqrt{3-T^{2}_{W}}},\\
t&=&\frac{S_{W}}{\sqrt{1-4S^{2}_{W}}}.
\end{eqnarray}
We mention that
$S_{W}\equiv\sin\theta_{W}$, $C_{W}\equiv
\cos\theta_{W}$, $T_{W}\equiv\tan\theta_{W}$
and $h_{W}=3-4S_{W}^{2}$
where $\theta_{W}$ is the Weinberg angle.
\section{Higgs Decays}
\label{sec:Higgs Decays}
~~~The main goal of our work is to study the signal strength of $h_1$ and the branching ratio of the other heavy scalar bosons $h_2$ and $h_3$ for different channels in the compact 341 model
\subsection{$h_{1}$ decay}
~~~The first purpose of this paper is to study the deviation
of our model from the Standard Model by calculating the signal strength $\mu$ of the lightest scalar boson $h_{1}$ and confront our results to the experimental data for each individual channels $b\bar{b}$, $WW^*$, $ZZ^*$, $\tau\bar{\tau}$ and $\gamma\gamma$.\\
\indent The signal strength of any process with a giving initial state i producing an Higgs h which decays to the final state f can be written as a product of the Higgs boson production cross section and its branching ratio in units of the corresponding value predicted by the SM \cite{Caetano}:
\begin{equation}
\mu_{xy}=\frac{\sigma_{341}(pp\longrightarrow h_{1})BR_{341}(h_{1}
	\longrightarrow xy)}{\sigma_{SM}(pp\longrightarrow h)BR_{SM}(h
	\longrightarrow xy)},
\end{equation}
where the superscript SM and 341 refer to the Standard Model and the compact 341 model respectively, while, x and y are any finale state, $\sigma_{i}$ and $\text{BR}_{i}$ (i=341, SM) are the corresponding cross section production taking gluon-gluon fusion (ggF) is the dominant contribution of the Higgs production and the branching ratio respectively. \indent In the compact 341 model, 
the coupling $tth_{1}$ is the same as tth of the Standard Model (SM), hence,
the cross section of the light Higgs $h_{1}$ production process is the same as the SM at the LHC:
\begin{equation}
\sigma_{341}(pp\longrightarrow h_{1})=\sigma_{SM}(pp\longrightarrow h),
\end{equation}
thus, the signal strength $\mu_{xy}$ becomes the ratio between the branching ratio of the SM and of the compact 341 model:
\begin{equation}
\mu_{xy}=\frac{BR_{341}}{BR_{SM}}=\frac{\Gamma_{SM}(h\longrightarrow
	all)\Gamma_{341}(h_{i}\longrightarrow
	xy)}{\Gamma_{341}(h_{i}\longrightarrow
	all)\Gamma_{SM}(h\longrightarrow xy)},
\end{equation}
 where the total decay width $\Gamma_{341}(h_1\longrightarrow
 all)$ turns out to be the same as the one of the SM Higgs boson $\Gamma_{SM}(h\longrightarrow
 all)$:
\begin{eqnarray}
\Gamma_{341}(h_1\longrightarrow all)&=&\Gamma_{341}(h_1\longrightarrow
b\overline{b})+\Gamma_{341}(h_1\longrightarrow
\tau^{+}\tau^{-})+\Gamma_{341}(h_1\longrightarrow WW^{*})
\nonumber\\&+&\Gamma_{341}(h_1\longrightarrow
ZZ^{*})+\Gamma_{341}(h_1\longrightarrow
\gamma\gamma)+\Gamma_{341}(h_1\longrightarrow \gamma
Z)\nonumber\\&+&\Gamma_{341}(h_1\longrightarrow gg).
\end{eqnarray}
\indent All the tree level couplings between the SM-like Higgs boson $h_1$ and all the fermions and the W boson, are the same as the SM, that makes the partial decay widths of $h_{1}$ into SM particles such as $b\bar{b}$, $\tau^{+}\tau^{-}$, $WW^{*}$ and gg have the same expressions as the SM \cite{Mebarki:2019eeh}, whereas, the partial decay width of $h_{1}$ into $ZZ^{*}$ is given by:
\begin{eqnarray}
\Gamma_{341}(h_{1}\longrightarrow Z^{*}Z)&=&
\frac{g^{4}m_{h_{1}}}{2048\pi^{3}}G^{2}_{h_{1}ZZ}F\bigg(\frac{m_{Z}}{m_{h_{1}}}\bigg)\bigg(\sum_{j\equiv u,d,c,s,b}(g_{jV}^{2}+g_{jA}^{2})
\nonumber\\&+&\sum_{\ell\equiv \text{leptons}}(g_{\ell V}^{2}+g_{\ell A}^{2})\bigg),
\end{eqnarray}
with $G_{h_{1}ZZ}=4g_{h_{1}ZZ}/(g^{2}\upsilon_{\rho})$ where $g_{ZZh_{1}}$ represents the coupling between a pair of Z bosons and $h_{1}$ (see table \ref{tab:i8}) and
\begin{eqnarray}
F(x)&=&-|1-x^{2}|(\frac{47}{2}x^{2}-\frac{13}{2}+\frac{1}{x^{2}})
-\frac{3}{2}(1-6x^{2}+4x^{4})\ln(x)\nonumber\\
&+&\frac{3(1-8x^{2}+20x^{4})}{\sqrt{4x^{2}-1}}\arccos(\frac{3x^{2}-1}{2x^{3}}),
\end{eqnarray}
with the parameter $x=\frac{m_{Z}}{m_{h_{1}}}$.\\
\indent The compact 341 model contains new charged particles including fermions, scalars (the new singly and doubly charged scalar bosons) and gauge bosons that will contribute to the decay amplitudes of the decays $h_{1}\longrightarrow \gamma\gamma$ and  $h_{1}\longrightarrow \gamma Z$ at one loop level. In the case of $h_1$, besides the contribution of $W^{\mp}$ boson and the top quark, the processes $\Gamma(h_{1}\longrightarrow \gamma\gamma (\gamma Z))$ receive new contributions that come from new particles (gauge and scalar bosons) such as $V^{\mp\mp}$, $K_{1}^{\mp}$, $h_{1}^{\mp}$, $h_{2}^{\mp}$ and $h^{\mp\mp}$, Furthermore, in our model,
the SM-like Higgs boson $h_{1}$ does not couple to the new heavy charged fermions, hence, the exotic quarks do not contribute to the one-loop decay amplitudes of the processes $h_{2,3}\longrightarrow(\gamma\gamma, Z\gamma$, gg). Thus, the partial decay widths of $h_{1}$ into $\gamma\gamma$ and Z$\gamma$ are given by \cite{Carena:2012xa,Mebarki:2019eeh}:
\begin{eqnarray}\label{eq:7}
\Gamma(h_{1}\longrightarrow
\gamma\gamma)&=&\frac{\alpha^{2}m_{h_{1}}^{3}}{1024\pi^{3}}\bigg|\sum_{V}\frac{g_{h_{1}VV}}{m^{2}_{V}}Q_{V}^{2}\mathcal{A}_{1}(\tau_{V})+\sum_{f}\frac{2g_{h_{1}ff}}{m_{f}}Q_{f}^{2}N_{c,f}\mathcal{A}_{\frac{1}{2}}(\tau_{f})\nonumber
\\&+&\sum_{f}\frac{g_{h_{1}SS}}{m_{S}^{2}}Q_{S}^{2}N_{c,S}\mathcal{A}_{0}(\tau_{S})\bigg|^{2},\\
\Gamma(h_{1}\longrightarrow \gamma
Z)&=&\frac{\alpha^{2}m_{h_{1}}^{3}}{512\pi^{3}}\bigg(1-\frac{M^{2}_{Z}}{M^{2}_{h_{1}}}\bigg)^{3}\bigg|\frac{2}{\upsilon}\frac{\mathcal{A}_{SM}}{\sin\theta_{W}}+\mathcal{A}\bigg|^{2},
\end{eqnarray}
where the symbol V, f and S refer to Spin 1, Spin $\frac{1}{2}$ and Spin 0
particles respectively, $Q_{V}, Q_{f}$, $Q_{S}$ are electric charges
of the vectors, fermions and scalars,
$g_{h_{1}SS}$,$g_{h_{1}VV}$ and $g_{h_{1}ff}$ represent the couplings
of the Higgs with scalars S, gauge bosons V and fermions f, $N_{c,f}, N_{c,S}$ are the number of fermion and scalar colors and the loop functions for V,f and S particles
$\mathcal{A}_{1}(\tau_{V})$,
$\mathcal{A}_{\frac{1}{2}}(\tau_{f})$ and
$\mathcal{A}_{0}(\tau_{S})$
are given by \cite{Carena:2012xa,Mebarki:2019eeh}:
\begin{eqnarray}
A_{1}(x)&=&-x^{2}\bigg(2x^{-2}+3x^{-1}+3(2x^{-1}-1)f(x^{-1})\bigg),\nonumber\\
A_{\frac{1}{2}}(x)&=&2x^{2}\bigg(x^{-1}+(x^{-1}-1)f(x^{-1})\bigg),\nonumber\\
A_{0}(x)&=&-x^{2}\bigg(x^{-1}-f(x^{-1})\bigg).
\end{eqnarray}
\indent The parameter $\tau_{i}=\frac{4m_{i}^{2}}{m^{2}_{h_{i}}}$ where i represents the corresponding particles in the loop and \cite{Mebarki:2019eeh}:
\begin{equation}
f(x)=\label{1-1} \left\lbrace
\begin{array}{ll}\bigskip
\arcsin^{2}\sqrt{x}\qquad \text{for}\qquad x\geq 1,\\
\frac{-1}{4}\bigg(\ln(\frac{1+\sqrt{1-x^{-1}}}{1-\sqrt{1-x^{-1}}
})-\imath\pi\bigg)^{2}\qquad \text{for}\qquad x< 1.
\end{array}
\right.
\end{equation}
\indent The factors $\mathcal{A}$ and $\mathcal{A}_{SM}$ represent the contributions of the new particles predicted by our
model and the contributions coming from the particles of the Standard Model respectively, their expressions are given by:
\begin{eqnarray}
\mathcal{A}&=&\frac{g_{h_{1}VV}}{m^{2}_{V}}g_{ZVV}\mathcal{A}_{1}(\tau_{V},\lambda_{V})+\widetilde{N}_{c,f}\frac{2g_{h_{1}ff}}{m_{f}}2Q_{f}(g_{Z\ell \ell}
+g_{Zrr})\mathcal{A}_{\frac{1}{2}}(\tau_{f},\lambda_{f})\nonumber\\
&-&\widetilde{N}_{c,s}\frac{2g_{h_{1}SS}}{m^{2}_{S}}Q_{S}g_{ZSS}\mathcal{A}_{0}(\tau_{S},\lambda_{S}),\\
\mathcal{A}_{SM}&=&\cos\theta_{W}\mathcal{A}_{1}(\tau_{W},\lambda_{W})+N_{C}\frac{Q_{t}(2T_{3}^{t}-4Q_{t}\sin\theta_{W}^{2})}{\cos\theta_{W}}\mathcal{A}_{\frac{1}{2}}(\tau_{t},\lambda_{t}),
\end{eqnarray}
Where $\lambda_{i}=\frac{4m_{i}^{2}}{m^{2}_{Z^{2}}}$, $T_{3}=\frac{1}{2}$ is the weak isospin of the top quark and
$\mathcal{A}_{i}(x,y)$ are the loop functions \cite{Carena:2012xa,Mebarki:2019eeh}:
\begin{eqnarray}
A_{1}(x,y)&=&4(3-\tan^{2}\theta_{W})I_{2}(x,y)+\bigg((1+2x^{-1})\tan^{2}\theta_{W}-(5+2x^{-1})\bigg)I_{1}(x,y),\nonumber\\
A_{\frac{1}{2}}(x,y)&=&I_{1}(x,y)-I_{2}(x,y),\nonumber\\
A_{0}(x,y)&=&I_{1}(x,y),
\end{eqnarray}
where:
\begin{eqnarray}
I_{1}(x,y)&=&\frac{xy}{2(x-y)}+\frac{x^{2}y^{2}}{2(x-y)^{2}}\bigg(f(x^{-1})-f(y^{-1})\bigg)+\frac{x^{2}y}{(x-y)^{2}}\bigg(g(x^{-1})-g(y^{-1})\bigg),\nonumber\\
I_{2}(x,y)&=&\frac{-xy}{2(x-y)}\bigg(f(x^{-1})-f(y^{-1})\bigg),
\end{eqnarray}
with
\begin{equation}
g(x)=\label{1-1} \left\lbrace
\begin{array}{ll}\bigskip
\sqrt{x^{-1}-1}\arcsin\sqrt{x}\qquad \text{for}\qquad x\geq 1\\
\frac{\sqrt{1-x^{-1}}}{2}\bigg(\ln(\frac{1+\sqrt{1-x^{-1}}}{1-\sqrt{1-x^{-1}}
})-\imath\pi\bigg)\qquad \text{for}\qquad x< 1
\end{array}
\right.
\end{equation}
where the couplings
$g_{h_1VV}$,$g_{ZVV}$,$g_{h_1ff}$,$g_{Zff}$,$g_{h_1SS}$ and $g_{ZSS}$ are
given in tables \ref{tab:i8},\ref{tab:i11},\ref{tab:i12}.
\subsection{The decay of the neutral heavy scalar bosons $h_{2}$ and $h_{3}$}
~~~For the neutral heavy scalar bosons $h_{2}$ and $h_{3}$, we calculate their branching ratios (BRs) where:
\begin{equation}
\text{BR}=\frac{\Gamma_{i}}{\Gamma(h\longrightarrow
	all)},
\end{equation}
with $\Gamma(h\longrightarrow
all)=\sum \Gamma_{i}$ is the total decay rate that represents the sum of the individual decay rates. The total decay width of $h_{2}$ is determined by the following channels:
\begin{eqnarray}
\Gamma_{341}(h_{2}\longrightarrow
all)&=&\Gamma_{341}(h_{2}\longrightarrow
\tau^+\tau^-,b\bar{b})+\Gamma_{341}(h_{2}\longrightarrow
ZZ)\nonumber\\
&+&\Gamma_{341}(h_{2}\longrightarrow
\gamma Z)+\Gamma_{341}(h_{2}\longrightarrow
\gamma\gamma)+\Gamma_{341}(h_{2}\longrightarrow
gg)\nonumber\\&+&\Gamma_{341}(h_{2}\longrightarrow h_{1}h_{1}).
\end{eqnarray}
\indent Furthermore, $h_{2}$ can also decay to new charged and neutral particles F including fermions, scalar and gauge bosons, if kinematically allowed which would require $m_{h_{2}}>2m_{F}$, therefore, in this case, $\Gamma_{341}(h_{2}\longrightarrow
all)$ has new contributions $\Gamma_{341}(h_{2}\longrightarrow
F)$ where F can be $K_{1}^{\mp}$, $K^{\prime0}$, $X^{\mp}$, $Y^{\mp}$, $V^{\mp\mp}$, $Z^{\prime}$ or exotic quarks.\\
\indent In our model,
the heavy scalar bosons $h_{2,3}$ does not couple to the SM top quark, therefore, the fermion contributions come only from exotic quarks  in the loop processes $\Gamma(h_{2}\longrightarrow gg,
\gamma\gamma, \gamma Z)$, this result in:
\begin{equation}
\Gamma(h_{2}\longrightarrow
gg)=\frac{\alpha^{2}_{S}M^{3}_{h_{2}}}{128\pi^{3}\upsilon_{\eta}^{2}}\alpha^{2}\bigg|\sum_{i=1}^{3}A_{\frac{1}{2}}(\tau_U)\bigg|^{2}+
\frac{\alpha^{2}_{S}M^{3}_{h_{2}}}{128\pi^{3}\upsilon_{\chi}^{2}}\gamma^{2}\bigg|\sum_{i=1}^{3}A_{\frac{1}{2}}(\tau_J)\bigg|^{2},
\end{equation}
where $\alpha_{S}$ is the strong coupling and $A_{\frac{1}{2}}(\tau_f)$ represents the loop function of fermions. \indent The one loop expressions for $h_{2,3}$ decays into final states including massless bosons $\gamma\gamma$ and $Z\gamma$ can be mediated by new contributions that come from the new charged particles namely $K_{1}^{\mp}$, $K^{\prime0}$, $X^{\mp}$, $Y^{\mp}$, $V^{\mp\mp}$, U, J, $h_{1}^{\mp}$, $h_{2}^{\mp}$ and $h^{\mp\mp}$, we note that the contribution of the $W^{\mp}$ boson is not included in these amplitudes, since there is no direct coupling between $h_{2,3}$ and the $W^{\mp}$ boson.\\
\indent Moreover, it is worth pointing out that the fermions and bosonic decays are similar to the $h_{1}$ case just one needs to
replace the couplings of table \ref{tab:i8} by those of table \ref{tab:i9} for $h_{2}$ and by table \ref{tab:i10} for $h_{3}$. \\
\indent Furthermore, the partial decay widths of $h_{2}$ into a pair of Higgs like-boson $h_1$, into a pair of exotic quarks (U and J), into a pair of gauge bosons $V_i$ ($V_i\equiv X^{\mp},V^{\mp\mp},K^{\prime0},K_1^{\mp},Y^{\mp}$) and into a pair of Z and $Z^{\prime}$ are given respectively by:
\begin{eqnarray}
\Gamma_{341}(h_{2}\longrightarrow h_{1}h_{1})&=&\frac{1}{16\pi
	m_{h_{2}}}(g_{h_{2}h_{1}h_{1}})^{2}\bigg(1-\frac{4m_{h_{1}}^{2}}{m^{2}_{h_{2}}}\bigg)^{\frac{1}{2}},\\
\Gamma_{341}(h_{2}\longrightarrow
UU)&=&\frac{3g^{2}}{32\pi}\frac{m_{U}^{2}m_{h_{2}}}{m_{W}^{2}}\bigg(1-\frac{4m^{2}_{U}}{m^{2}_{h_{3}}}\bigg)^{\frac{3}{2}}\bigg(\frac{\upsilon_{\rho}}{\upsilon_{\eta}}\alpha\bigg)^{2},\\
\Gamma_{341}(h_{2}\longrightarrow
JJ)&=&\frac{3g^{2}}{32\pi}\frac{m_{J}^{2}m_{h_{2}}}{m_{W}^{2}}\bigg(1-\frac{4m^{2}_{J}}{m^{2}_{h_{2}}}\bigg)^{\frac{3}{2}}\bigg(\frac{\upsilon_{\rho}}{\upsilon_{\chi}}\gamma\bigg)^{2},\\
\Gamma_{341}(h_{2}\longrightarrow V_iV_i)&=&\frac{k_{i}}{4\pi}\frac{M_{W}^{4}}{M_{h_{2}}\upsilon_{\rho}^2}\bigg(1-\frac{4M_{V_i}^{2}}{M_{h_{2}}^{2}}\bigg)^{\frac{1}{2}}\bigg(3+\frac{1}{4}\frac{M_{h_{2}}^{4}}{M^{4}_{V_i}}-\frac{M_{h_{2}}^{2}}{M^{2}_{V_i}}\bigg)\label{eq:doudi},\\
\Gamma(h_{2}\longrightarrow
ZZ)&=&\frac{1}{8\pi}\frac{M_{Z}^{4}}{M_{h_{2}}\upsilon_{\rho}^{2}}\bigg(1-\frac{4M^{2}_{Z}}{M^{2}_{h_{2}}}\bigg)^{\frac{1}{2}}\bigg(3+\frac{M^{4}_{h_{2}}}{M_{Z}^{4}}-\frac{M^{2}_{h_{2}}}{M_{Z}^{2}}\bigg)G^{2}_{ZZh_{2}},\label{eq:ZBo}
\end{eqnarray}
\begin{eqnarray}
\Gamma(h_{2}\longrightarrow
Z^{\prime}Z^{\prime})&=&\frac{1}{8\pi}\frac{M_{Z}^{4}}{M_{h_{2}}\upsilon_{\rho}^{2}}\bigg(1-\frac{4M^{2}_{Z^{\prime}}}{M^{2}_{h_{2}}}\bigg)^{\frac{1}{2}}\bigg(3+\frac{M^{4}_{h_{2}}}{M_{Z^{\prime}}^{4}}-\frac{M^{2}_{h_{2}}}{M_{Z^{\prime}}^{2}}\bigg)G^{2}_{Z^{\prime}Z^{\prime}h_{2}}\label{eq:ZBo1}.
\end{eqnarray}
where the expression of the trilinear coupling $h_{2}h_{1}h_{1}$ 
is given in table \ref{tab:i9}, $m_{U}$, $m_{J}$ are the exotic quarks masses, $\upsilon_{\eta}$ is the vacuum expectation value and the $k_i$ coefficients are given in the following table: 
\begin{table}[H]
	\tbl{\label{tab:i11k}$k_i$ coefficients.}
	{\begin{tabular}{@{}ccccccccccccccc@{}} \toprule
		Higgs&$k_{X^{\mp}}$&$k_{V^{\mp\mp}}$&$k_{K^{\prime0}}$&$k_{K_{1}^{\mp}}$&$k_{Y^{\mp}}$\\ \colrule
		$h_2$&$\frac{\upsilon_{\chi}}{\upsilon_{\rho}}\gamma$&$\frac{\upsilon_{\chi}}{\upsilon_{\rho}}\gamma$&$\frac{\upsilon_{\eta}}{\upsilon_{\rho}}\alpha$&$\frac{\upsilon_{\eta}}{\upsilon_{\rho}}\alpha$&$\frac{1}{\upsilon_{\rho}}(\upsilon_{\chi}\gamma+\upsilon_{\eta}\alpha)$\\
			\botrule
	\end{tabular}}
\end{table}
 while the expressions of $G_{ZZh_{2}}$ and $G_{Z^{\prime}Z^{\prime}h_{2}}$ are given by:
\begin{eqnarray}
G_{ZZh_{2}}&=&\frac{C_{W}^{2}}{2}\bigg(\frac{3}{2}\frac{\upsilon_{\chi}}{\upsilon_{\rho}}(C^{23})^{2}\gamma+4(C^{24})^{2}\frac{S_{W}^{2}}{1-4S_{W}^{2}}\frac{\upsilon_{\chi}}{\upsilon_{\rho}}\gamma+\frac{12}{\sqrt{6}}
\frac{S_{W}}{\sqrt{1-4S_{W}^{2}}}\frac{\upsilon_{\chi}}{\upsilon_{\rho}}C^{23}C^{24}\gamma\nonumber\\&+&\frac{4}{3}(C^{22})^{2}\frac{\upsilon_{\eta}}{\upsilon_{\rho}}\alpha
+\frac{(C^{23})^{2}}{6}\frac{\upsilon_{\eta}}{\upsilon_{\rho}}\alpha-\frac{4}{\sqrt{18}}C^{22}C^{23}\frac{\upsilon_{\eta}}{\upsilon_{\rho}}\alpha\bigg),\\
G_{Z^{\prime}Z^{\prime}h_{2}}&=&\frac{C_{W}^{2}}{2}\bigg(\frac{3}{2}\frac{\upsilon_{\chi}}{\upsilon_{\rho}}(C^{33})^{2}\gamma+4(C^{34})^{2}\frac{S_{W}^{2}}{1-4S_{W}^{2}}\frac{\upsilon_{\chi}}{\upsilon_{\rho}}\gamma+\frac{12}{\sqrt{6}}
\frac{S_{W}}{\sqrt{1-4S_{W}^{2}}}\frac{\upsilon_{\chi}}{\upsilon_{\rho}}C^{33}C^{34}\gamma\nonumber\\&+&\frac{4}{3}(C^{32})^{2}\frac{\upsilon_{\eta}}{\upsilon_{\rho}}\alpha
+\frac{(C^{33})^{2}}{6}\frac{\upsilon_{\eta}}{\upsilon_{\rho}}\alpha-\frac{4}{\sqrt{18}}C^{32}C^{33}\frac{\upsilon_{\eta}}{\upsilon_{\rho}}\alpha\bigg).
\end{eqnarray}
\indent Regarding, the third neutral scalar boson $h_{3}$, its total decay width is composed by the following decays: 
\begin{eqnarray}
\Gamma_{341}(h_{3}\longrightarrow
all)&=&\Gamma_{341}(h_{3}\longrightarrow
\tau^+\tau^-, b\bar{b})+\Gamma_{341}(h_{3}\longrightarrow \text{exotic~quarks})\nonumber\\
&+&\Gamma_{341}(h_{3}\longrightarrow
\gamma\gamma)+\Gamma_{341}(h_{3}\longrightarrow
Z\gamma)+\Gamma_{341}(h_{3}\longrightarrow
gg)\nonumber\\
&+&\Gamma_{341}(h_{3}\longrightarrow h_{2}h_{2})
+\Gamma_{341}(h_{3}\longrightarrow h_{1}h_{1})\nonumber\\
&+&\Gamma_{341}(h_{3}\longrightarrow
h_{2}h_{1})+\Gamma_{341}(h_{3}\longrightarrow VV),
\end{eqnarray}
where V represents the gauge bosons $X^{\pm}$, $V^{\pm\pm}$, $K^{\prime0}$, $K_{1}^{\pm}$, $Y^{\mp}$, Z and $Z^{\prime}$.
It is worth to point out that $h_{3}$ may decay into a pair of $h_{1}h_{1}$, $h_{2}h_{2}$ and $h_{1}h_{2}$ where their decay widths are given by:
\begin{eqnarray}
\Gamma_{341}(h_{3}\longrightarrow h_{1}h_{1})&=&\frac{1}{16\pi
	m_{h_{3}}}(g_{h_{3}h_{1}h_{1}})^{2}\bigg(1-\frac{4m_{h_{1}}^{2}}{m^{2}_{h_{3}}}\bigg)^{\frac{1}{2}},\\
\Gamma_{341}(h_{3}\longrightarrow h_{2}h_{2})&=&\frac{1}{16\pi
	m_{h_{3}}}(g_{h_{3}h_{2}h_{2}})^{2}\bigg(1-\frac{4m_{h_{2}}^{2}}{m^{2}_{h_{3}}}\bigg)^{\frac{1}{2}},\\
\Gamma_{341}(h_{3}\longrightarrow h_{2}h_{1})&=&\frac{1}{16\pi
	m_{h_{3}}^{2}}(g_{h_{3}h_{2}h_{1}})^{2}\bigg(\frac{m_{h_{2}}^{4}}{m_{h_{3}}^{2}}-\frac{2m_{h_{1}}^{2}m_{h_{2}}^{2}}{m^{2}_{h_{3}}}-2m_{h_{2}}^{2}+\frac{m_{h_{1}}^{4}}{m_{h_{3}}^{2}}-2m_{h_{1}}^{2}\nonumber\\&+&m_{h_{3}}^{2}\bigg)^{\frac{1}{2}},
\end{eqnarray}
where $g_{h_{3}h_{2}h_{1}}$, $g_{h_{3}h_{1}h_{1}}$ and $g_{h_{3}h_{2}h_{2}}$ represents trilinear terms $h_{3}h_{2}h_{1}$, $h_{3}h_{1}h_{1}$ and $h_{3}h_{2}h_{2}$ respectively, their expressions are given in table \ref{tab:i10}.\\
\indent The partial decay width of $h_{3}$ into a pair of gauge bosons $V_i$ is given by Eq \eqref{eq:doudi} where in the case of $h_{3}$, the coefficients $k_i$ are given by:
\begin{table}[H]
	\tbl{\label{tab:i11j}$k_i$ coefficients.}
	{\begin{tabular}{@{}ccccccccccccccc@{}} \toprule
			Higgs&$k_{X^{\mp}}$&$k_{V^{\mp\mp}}$&$k_{K^{\prime0}}$&$k_{K_{1}^{\mp}}$&$k_{Y^{\mp}}$\\ \colrule
			$h_3$&$\frac{\upsilon_{\chi}}{\upsilon_{\rho}}\sigma$&$\frac{\upsilon_{\chi}}{\upsilon_{\rho}}\sigma$&$\frac{\upsilon_{\eta}}{\upsilon_{\rho}}\beta$&$\frac{\upsilon_{\eta}}{\upsilon_{\rho}}\beta$&$\frac{1}{\upsilon_{\rho}}(\upsilon_{\chi}\sigma+\upsilon_{\eta}\beta)$\\
			\botrule
	\end{tabular}}
\end{table}
\indent The partial decay width of $h_3$ into a pair of SM Z and $Z^{\prime}$ gauge bosons are given by the Eqs \eqref{eq:ZBo} and \eqref{eq:ZBo1} respectively only one needs to replace $m_{h_{2}}$ with $m_{h_{3}}$ and both $G_{ZZh_{2}}$ and $G_{Z^{\prime}Z^{\prime}h_{2}}$ with $G_{ZZh_{3}}$ and $G_{Z^{\prime}Z^{\prime}h_{3}}$ respectively, where:
\begin{eqnarray}
G_{ZZh_{3}}&=&\frac{C^{2}_{W}}{2}\bigg(\frac{3}{2}\frac{\upsilon_{\chi}}{\upsilon_{\rho}}(C^{23})^{2}\sigma+4(C^{24})^{2}\frac{S_{W}^{2}}{1-4S_{W}^{2}}\frac{\upsilon_{\chi}}{\upsilon_{\rho}}\sigma+\frac{12}{\sqrt{6}}
\frac{S_{W}}{\sqrt{1-4S_{W}^{2}}}\frac{\upsilon_{\chi}}{\upsilon_{\rho}}C^{23}C^{24}\sigma\nonumber\\&+&\frac{4}{3}(C^{22})^{2}\frac{\upsilon_{\eta}}{\upsilon_{\rho}}\beta
+\frac{(C^{23})^{2}}{6}\frac{\upsilon_{\eta}}{\upsilon_{\rho}}\beta-\frac{4}{\sqrt{18}}C^{22}C^{23}\frac{\upsilon_{\eta}}{\upsilon_{\rho}}\beta\bigg),\\
G_{Z^{\prime}Z^{\prime}h_{3}}&=&\frac{C^{2}_{W}}{2}\bigg(\frac{3}{2}\frac{\upsilon_{\chi}}{\upsilon_{\rho}}(C^{33})^{2}\sigma+4(C^{34})^{2}\frac{S_{W}^{2}}{1-4S_{W}^{2}}\frac{\upsilon_{\chi}}{\upsilon_{\rho}}\sigma+\frac{12}{\sqrt{6}}
\frac{S_{W}}{\sqrt{1-4S_{W}^{2}}}\frac{\upsilon_{\chi}}{\upsilon_{\rho}}C^{33}C^{34}\sigma\nonumber\\&+&\frac{4}{3}(C^{32})^{2}\frac{\upsilon_{\eta}}{\upsilon_{\rho}}\beta
+\frac{(C^{33})^{2}}{6}\frac{\upsilon_{\eta}}{\upsilon_{\rho}}\beta-\frac{4}{\sqrt{18}}C^{32}C^{33}\frac{\upsilon_{\eta}}{\upsilon_{\rho}}\beta\bigg).
\end{eqnarray}
\indent The partial decay widths of the $h_{3}$ into gluons and into a pair of exotic quarks are given respectively by:
\begin{eqnarray}
\Gamma(h_{3}\longrightarrow
gg)&=&\frac{\alpha^{2}_{S}M^{3}_{h_{3}}}{128\pi^{3}\upsilon_{\eta}^{2}}\beta^{2}\bigg|\sum_{i=1}^{3}AU_{\frac{1}{2}}(\tau_U)\bigg|^{2}+
\frac{\alpha^{2}_{S}M^{3}_{h_{3}}}{128\pi^{3}\upsilon_{\chi}^{2}}\sigma^{2}\bigg|\sum_{i=1}^{3}AJ_{\frac{1}{2}}(\tau_J)\bigg|^{2},\\
\Gamma_{341}(h_{3}\longrightarrow
UU)&=&\frac{3g^{2}}{32\pi}\frac{m_{U}^{2}m_{h_{3}}}{m_{W}^{2}}\bigg(1-\frac{4m^{2}_{U}}{m^{2}_{h_{3}}}\bigg)^{\frac{3}{2}}\bigg(\frac{\upsilon_{\rho}}{\upsilon_{\eta}}\beta\bigg)^{2},\\
\Gamma_{341}(h_{3}\longrightarrow
JJ)&=&\frac{3g^{2}}{32\pi}\frac{m_{J}^{2}m_{h_{3}}}{m_{W}^{2}}\bigg(1-\frac{4m^{2}_{J}}{m^{2}_{h_{3}}}\bigg)^{\frac{3}{2}}\bigg(\frac{\upsilon_{\rho}}{\upsilon_{\chi}}\sigma\bigg)^{2}.
\end{eqnarray}
\section{Numerical Analysis}
\label{sec:Numerical Analysis}
~~~In this section, we discuss our numerical results of the signal strength of the Higgs
like-boson $h_{1}$ and the branching ration (BR) of the other scalars
$h_{2}$ and $h_{3}$. All the expressions of the compact 341 model
are related to $\upsilon_{\chi}$, $\upsilon_{\eta}$, $\upsilon_{\rho}$,
and to the scalar parameters $\lambda_{1..9}$.\\
\indent In our work, we consider the following scenario:\\
$\upsilon_{\rho}$=246 GeV, $\upsilon_{\chi}=\upsilon_{\eta}$=2 TeV and $m_{\text{exotic~quarks}}$=750 GeV \cite{Beringer}, while, for the scalar parameters $\lambda_{ 1..9}$, we make random choices where we constrained them using the theoretical constraints that we have discussed in the text. 
\subsection{The signal strength}
\begin{figure}[h]
	\includegraphics[width=.26\textwidth]{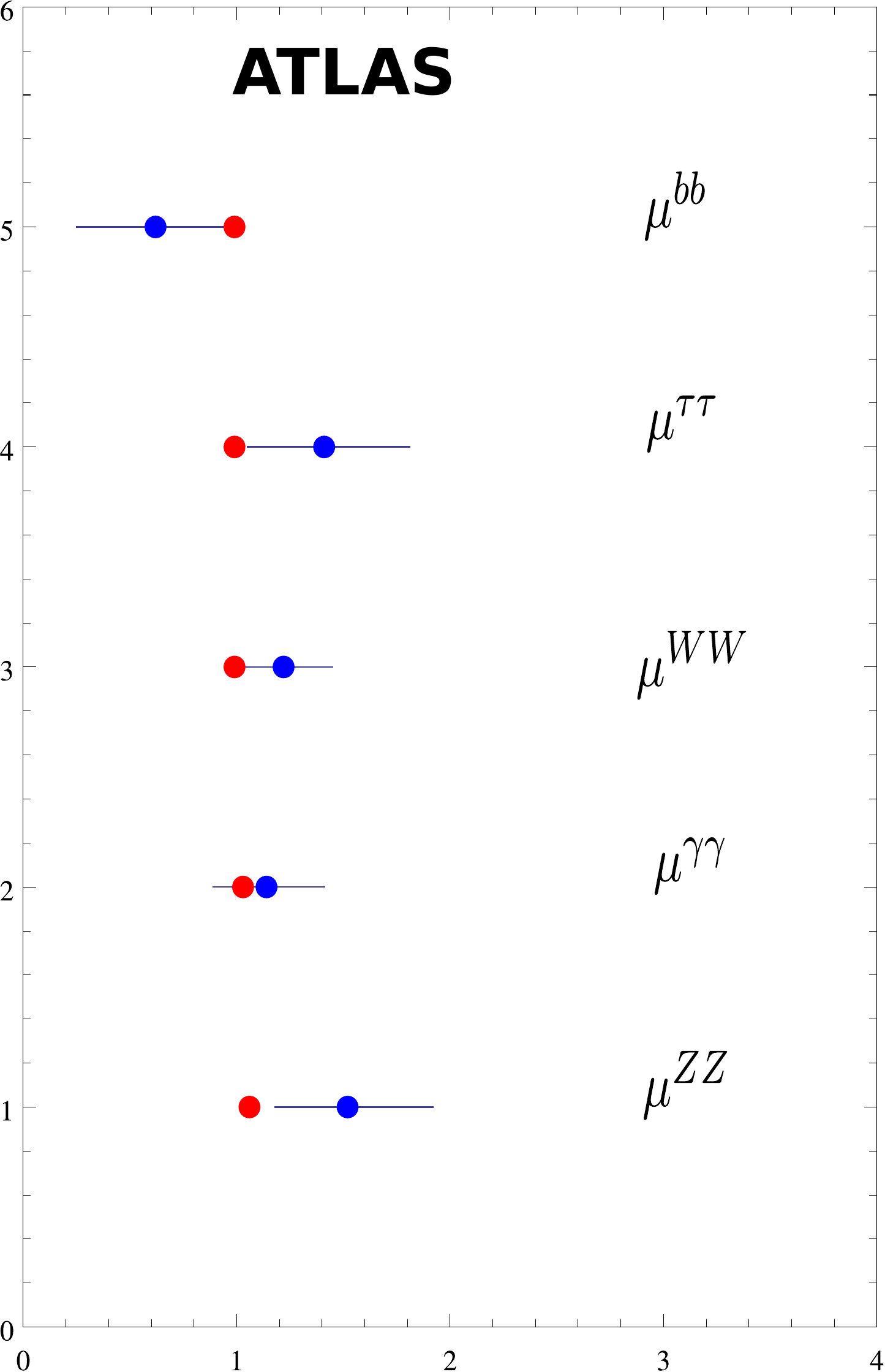}
	\hfill
	\includegraphics[width=.28\textwidth]{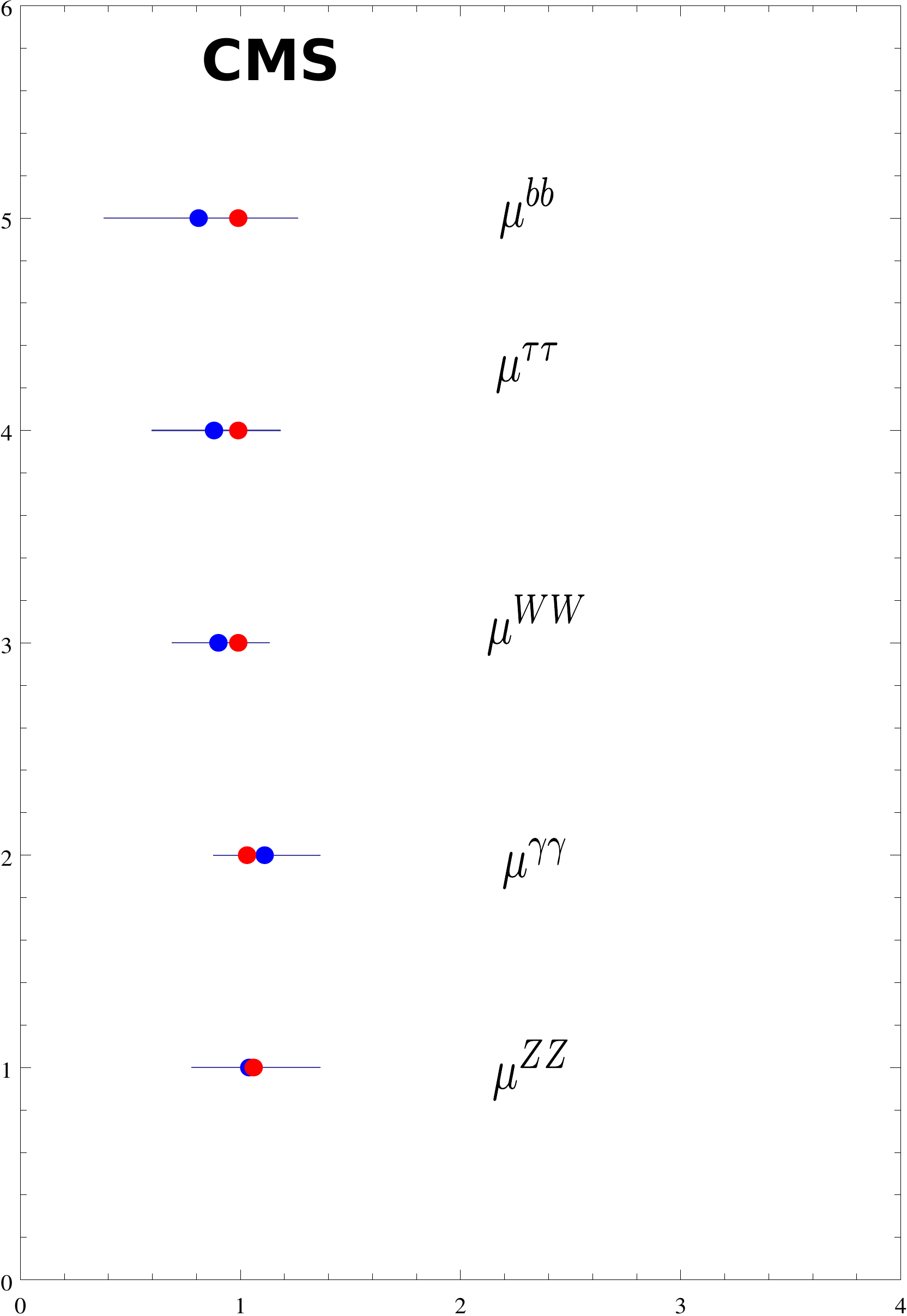}
	\hfill
	\includegraphics[width=.28\textwidth]{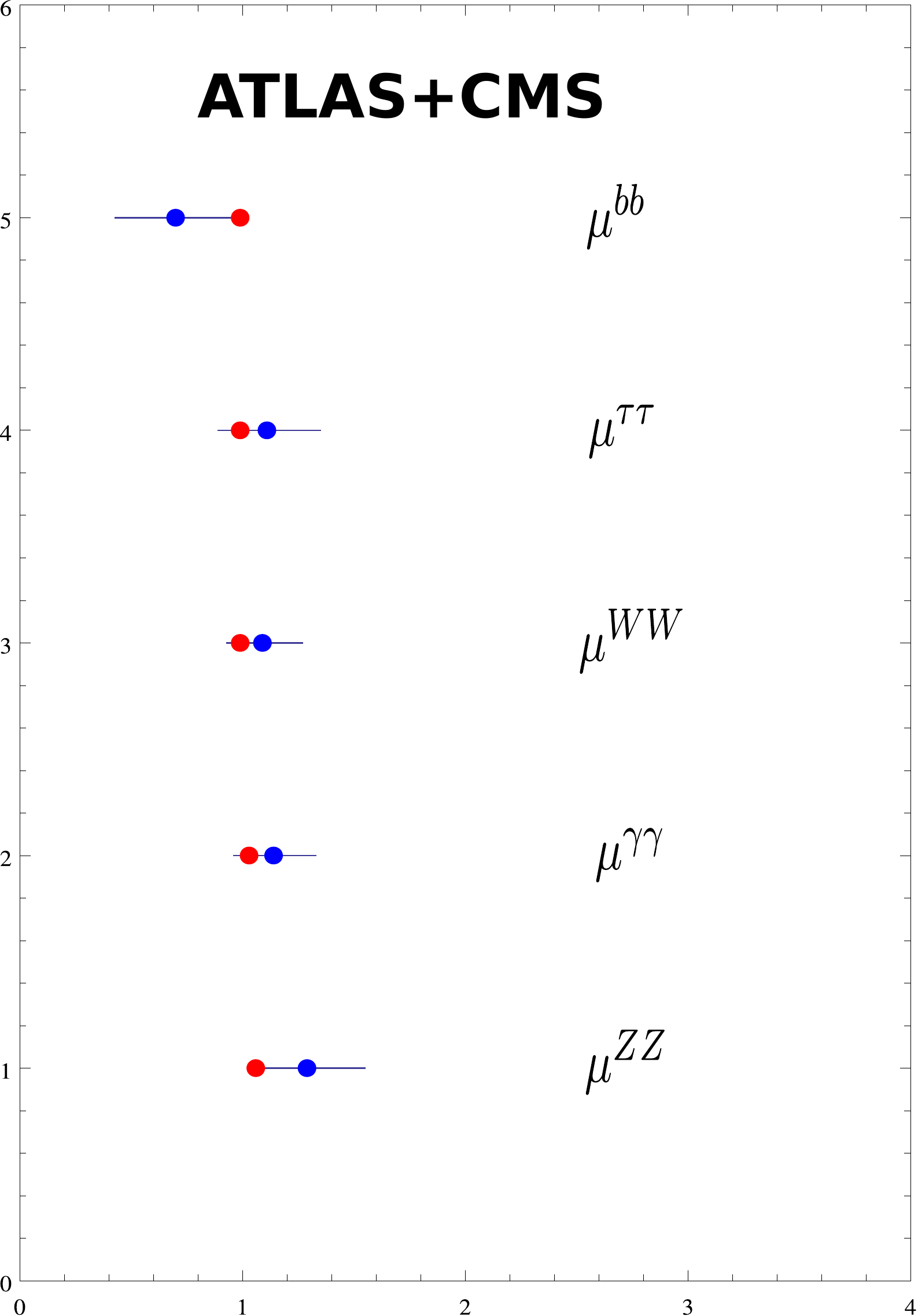}
	\caption{\label{fig:i} The signal strength results of $h_{1}$ for various decay modes of the compact 341 model (red points) and the experimental data (blue points)}
\end{figure}

~~~~Any extension of the SM must possess a scalar with 125 GeV of mass, in our model, we identify $h_{1}$ as the SM Higgs boson, therefore, the signal strength is discussed to clarify if the lightest scalar boson $h_{1}$ recovers the SM Higgs boson and to check the validity of our model.\\
\indent We calculated the signal strength of $h_{1}$ where we take into account that the most dominant production channel of $h_1$ comes from the gluon-gluon fusion, while, its decay is into several channels namely $b\bar{b}$, $WW^{*}$, $ZZ^{*}$, $\gamma\gamma$ and $\tau^{+}\tau^{-}$.\\
\indent The fermionic decays and the bosonic decays are possible at the tree-level, whereas we apply one-loop expressions for the decays into final states including massless
bosons (that is gg, $\gamma\gamma$ and Z$\gamma$).\\
\indent Table \ref{tab:1}, shows the results of our model and the available experimental signal strength values from LHC Run 1 for the combination of ATLAS and CMS, and separately for each experiment, for the combined $\sqrt{s}$ = 7 and 8 TeV data for different Higgs boson decay channels. 
 These results
are obtained assuming that the Higgs boson production process cross sections at s = 7 and 8 TeV are the same as
in the SM \cite{Khachatryan:2016vau}.\\
\indent Figure \ref{fig:i} shows the signal strengths for the various decay modes $b\bar{b}$, $WW^{*}$, $ZZ^{*}$, $\gamma\gamma$ and $\tau^{+}\tau^{-}$ in the compact 341 model with the data reported at ATLAS, CMS and the combined ATLAS+ CMS Run1. 
Our results can fit the current data withing the experimental errors which makes the compact 341 model in perfect agreement with the values measured by ATLAS and CMS. That ensures the viability of the compact 341 model to be an available model in the future work at the LHC.
\begin{table}[H]
\tbl{\label{tab:1} The SM Higgs boson signal strengths of the official ATLAS, CMS and ATLAS and CMS
	combination for Run 1 \cite{Khachatryan:2016vau}, based on 25 $\text{fb}^{-1}$ of integrated luminosity and of the compact 341 model.}
{\begin{tabular}{@{}ccccccccccccccc@{}} \toprule
Decay channel & ATLAS & CMS & ATLAS+CMS &  The compact 341 model\\ \colrule
		$\mu^{\gamma\gamma}$&$1.14^{+0.27}_{-0.25}$&$1.11^{+0.25}_{-0.23}$&$1.14^{+0.19}_{-0.18}$&1.03\\
		$\mu^{ZZ}$&$1.52^{+0.40}_{-0.34}$&$1.04^{+0.32}_{-0.26}$&$1.29^{+0.26}_{-0.23}$&1.06\\
		$\mu^{WW}$&$1.22^{+0.23}_{-0.21}$&$0.90^{+0.23}_{-0.21}$&$1.09^{+0.18}_{-0.16}$&0.99\\
		$\mu^{\tau\tau}$&$1.41^{+0.40}_{-0.36}$&$0.88^{+0.30}_{-0.28}$&$1.11^{+0.24}_{-0.22}$&0.99\\
		$\mu^{bb}$&$0.62^{+0.37}_{-0.37}$&$0.81^{+0.45}_{-0.43}$&$0.70^{+0.29}_{-0.27}$&0.99\\
			\botrule
	\end{tabular}}
\end{table}
\subsection{Searches for $h_{i}$ i=2,3}
\begin{figure}[H]
	\begin{center}
		\includegraphics[width=.58\textwidth]{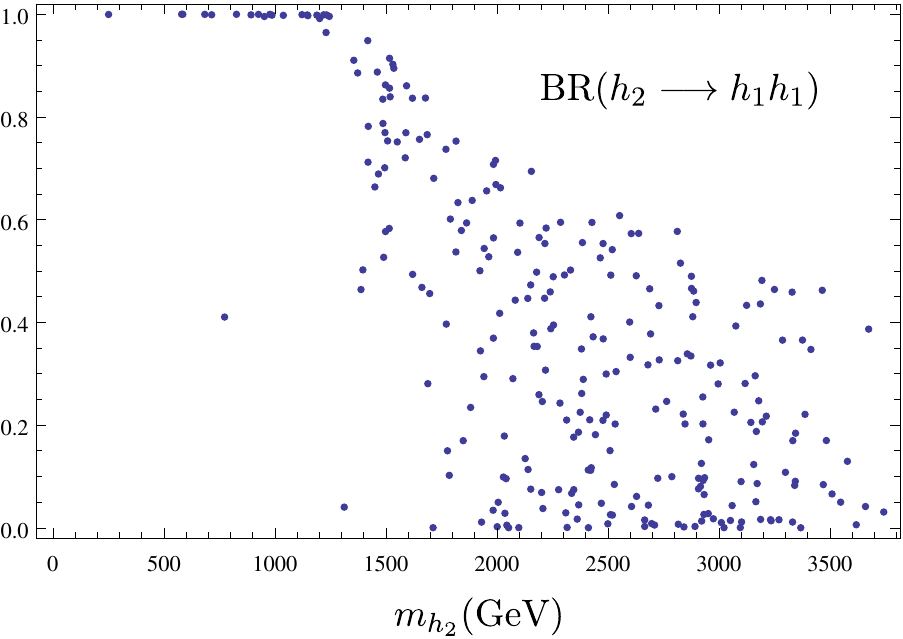}
		\caption{\label{fig:i5} The branching ratio of $h_{2}\longrightarrow h_{1}h_{1}$ versus the heavy Higgs $h_{2}$ mass.}
	\end{center}
\end{figure}

\begin{figure}[H]
	\begin{center}
			\includegraphics[width=.58\textwidth]{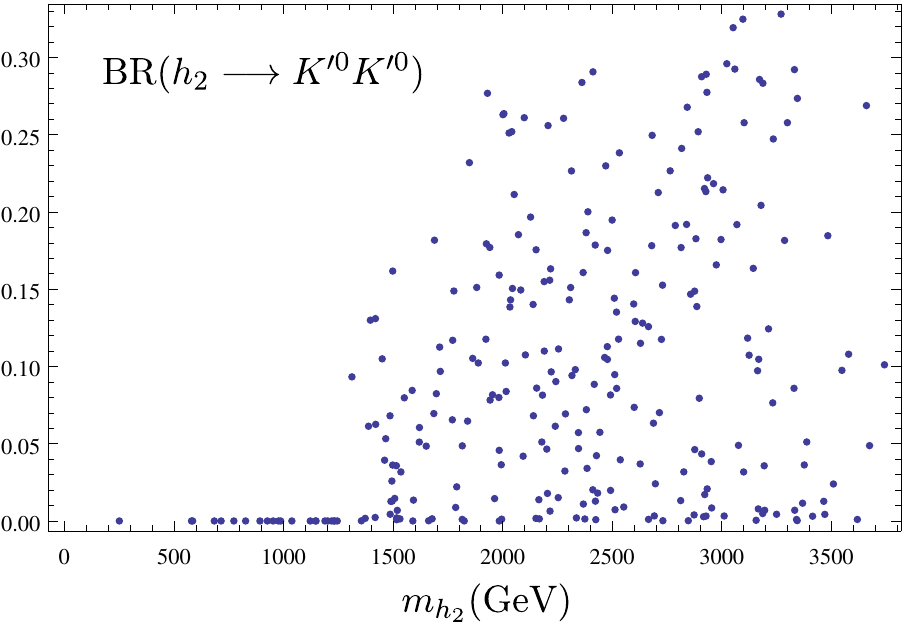}
				\caption{\label{fig:iVV} The branching ratio of  $h_{2}\longrightarrow K^{0\prime}K^{0\prime}$ versus the heavy Higgs $h_{2}$ mass.}
			\end{center}
		\end{figure}

\begin{figure}[H]
		\begin{center}
			\includegraphics[width=.58\textwidth]{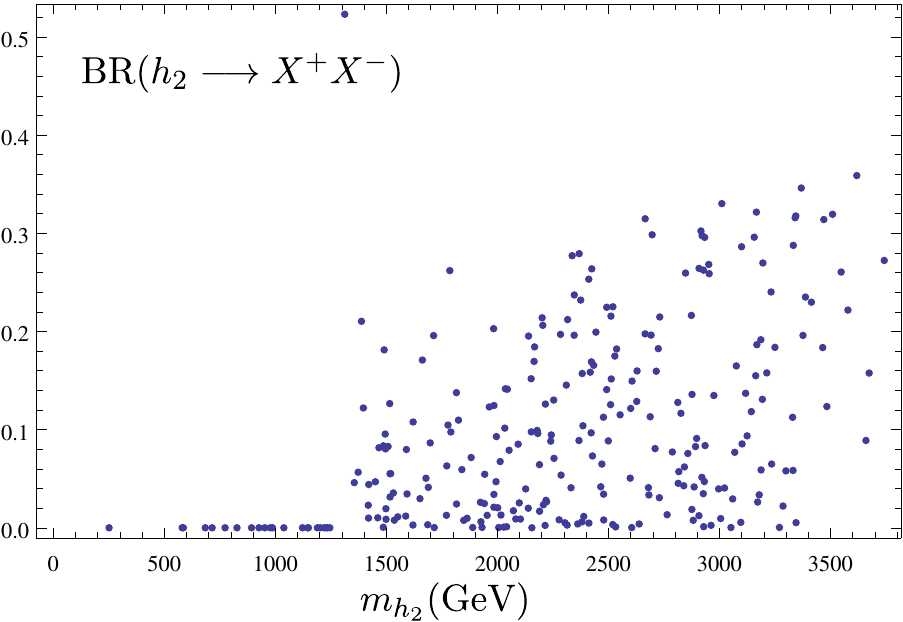}
			\hfill
			\includegraphics[width=.58\textwidth]{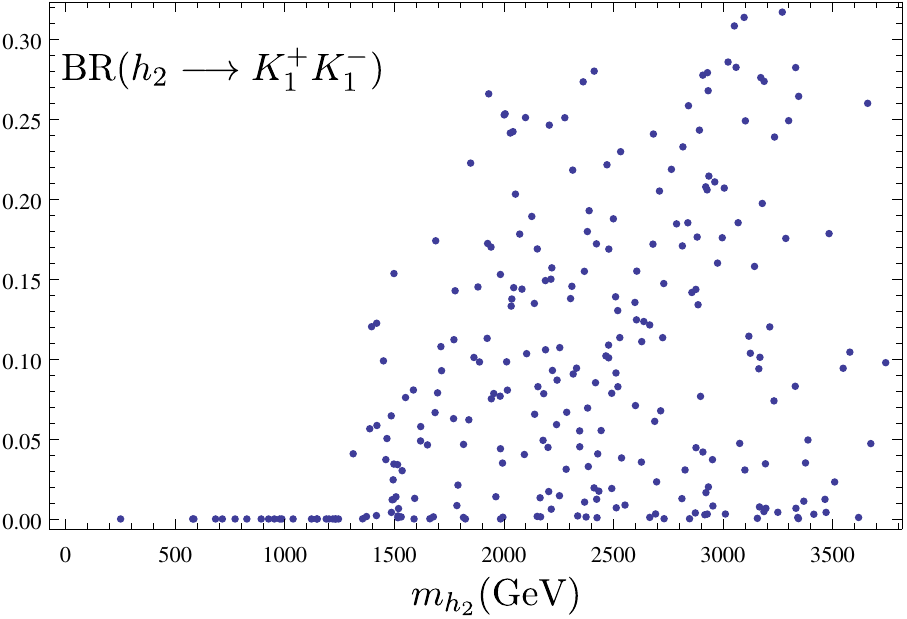}
			\hfill
			\includegraphics[width=.58\textwidth]{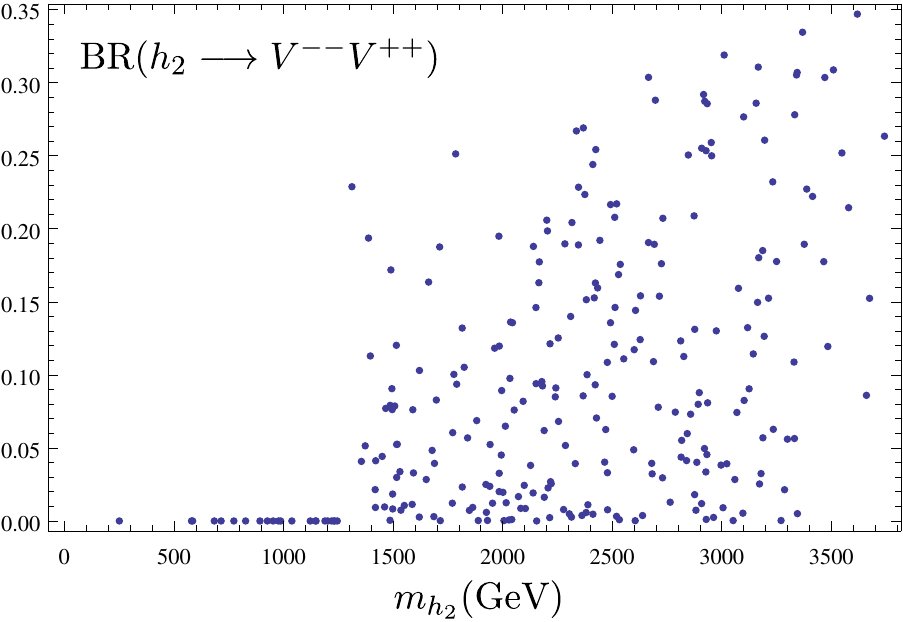}
		\caption{\label{fig:iVV1} The branching ratio of $h_{2}\longrightarrow X^+X^+$, $h_{2}\longrightarrow K_1^+K_1^-$ and $h_{2}\longrightarrow V^{++}V^{--}$ versus the heavy Higgs $h_{2}$ mass.}
	\end{center}
\end{figure}

~~~Figure \ref{fig:i5} shows BR$(h_{2}\longrightarrow h_{1}h_{1})$ as a function of the heavy Higgs $h_2$ mass for $v_\chi=v_\eta$=2 TeV where in the allowed parameter space, the mass of $m_{h_{2}}$ ranges from 500 GeV to 1.3 TeV. Its clear that in the region 500-1300 GeV , the decay channel $h_{2}\longrightarrow h_{1}h_{1}$ is the most important one with a BR $\geq 90\%$, this happens because the trilinear coupling $h_1h_2h_2$ increases about one order of magnitude $v_\eta$. Interestingly, one can has a new production mechanism
for $h_2$ namely $pp\longrightarrow h_{2}\longrightarrow h_{1}h_{1}$. This production channel might be useful to
increase the signal of the double Higgs production at the LHC. \\
\indent As the mass becomes larger than 1300 GeV,
 we can notice that the decay modes of $h_2$ into the heavy gauge bosons $K^{\prime0}$, $K_{1}^{\mp}$, $Y^{\mp}$, $V^{\mp\mp}$, $X^{\mp}$ and $Z^{\prime}$ and into exotic quarks are kinetically allowed, in this case, the total decay width contains additional channels $h_{2}\longrightarrow VV$ and $h_{2}\longrightarrow Q\overline{Q}$ where V$\equiv K^{\prime0}$, $K_{1}^{\mp}$, $Y^{\mp}$, $V^{\mp\mp}$, $X^{\mp}$ and $Z^{\prime}$ and Q$\equiv$ exotic quarks. The maximum value of the branching ratio of the process $h_2$ into gauge bosons $K^{\prime0}$, $K_{1}^{\mp}$, $V^{\mp\mp}$ and $X^{\mp}$ plateaus close to $\sim$0.35, $\sim$0.26 as we shown in tables \ref{tab:1n} and \ref{tab:12n} and in figures \ref{fig:iVV} and \ref{fig:iVV1}, therefore, one may notice that those decays are the most significant decay channel and the decay channel  $h_{2}\longrightarrow h_{1}h_{1}$ becomes weak and contributing roughly $20\%$.\\
 \indent The branching ratios of the remaining decay modes are tabulated in the following tables: 
\begin{table}[H]
	\tbl{\label{tab:1n} The branching ratios (BRs) of the heavy scalar $h_{2}$ for differnet channels.}
	{\begin{tabular}{@{}ccccccccccccccc@{}} \toprule
			Decay channel & BR where $m_{h_{2}}\in$ [500-1300](GeV) &  BR where $m_{h_{2}}\in$ [1300-3500](GeV)\\ \colrule
			$b\bar{b}$&$\sim \mathcal{O}(10^{-5})$&[$10^{-5}-10^{-6}$]\\
			$\tau^{+}\tau^{-}$&$\sim \mathcal{O}(10^{-7})$&$\sim 10^{-7}$\\
			$\gamma\gamma$&$\sim \mathcal{O}(10^{-7})$&$\sim 10^{-7}$\\
			$\gamma Z$&$\sim \mathcal{O}(10^{-3})$&[$10^{-3}-2.6.10^{-3}$]\\
			$gg$&$\sim \mathcal{O}(10^{-4})$&[$10^{-4}-10^{-5}$]\\
				ZZ& $\sim \mathcal{O}(10^{-32})$&[$10^{-32}-10^{-31}$]\\
			$Z^{\prime}Z^{\prime}$&$\sim \mathcal{O}(10^{-4})$&[$10^{-3}-10^{-2}$]\\
			$K^{\prime0}K^{\prime0}$	&0&[$10^{-3}$-0.26]\\
				$Y^{+}Y^{-}$&0&[$10^{-4}$-$10^{-2}$]\\
			$X^{+}X^{-}$&0&[$10^{-3}$-0.35]\\
			$V^{++}V^{--}$&0&[$10^{-3}$-0.34]\\
			\botrule
	\end{tabular}}
\end{table}

\begin{figure}[H]
	\begin{center}
		\includegraphics[width=.55\textwidth]{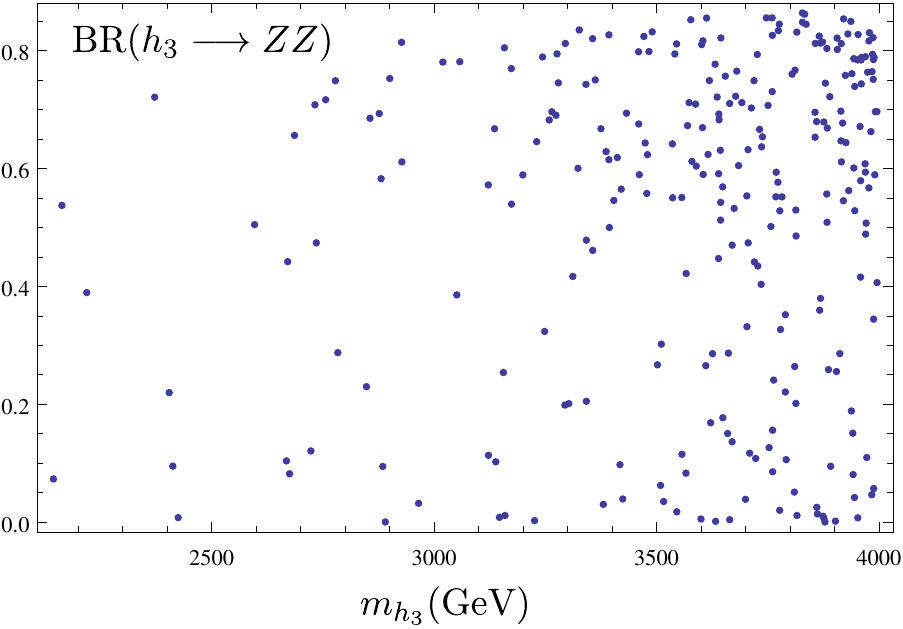}
		\caption{\label{fig:i6} The branching ratio of $h_{3}\longrightarrow ZZ$ as a function of the heavy Higgs $h_{3}$ mass.}
				\end{center}
		\end{figure}

\begin{figure}[h]
		\begin{center}
		\includegraphics[width=.55\textwidth]{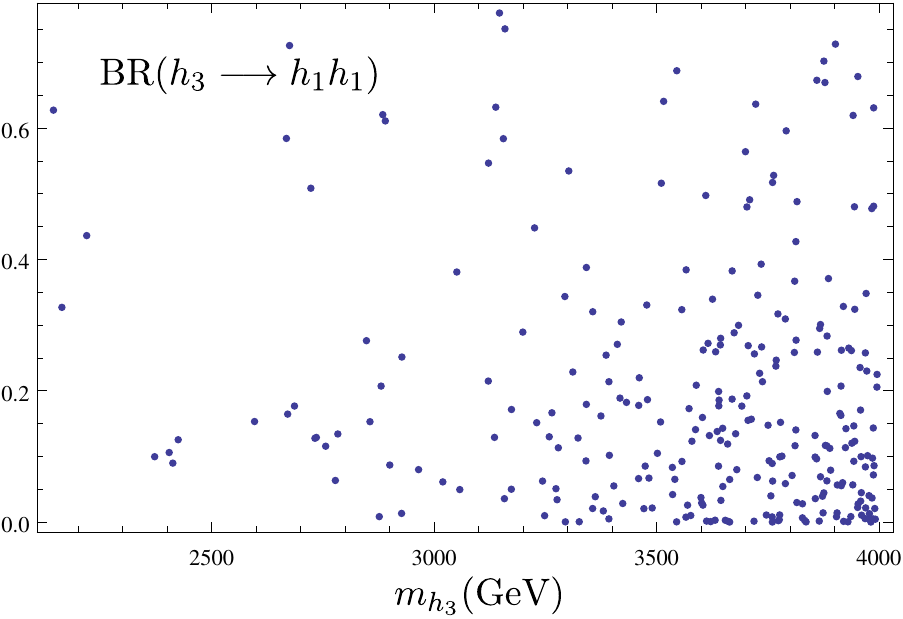}
		\caption{\label{fig:h11} The branching ratio of  $h_{3}\longrightarrow h_1h_1$ as a function of the heavy Higgs $h_{3}$ mass.}
	\end{center}
\end{figure}

\begin{figure}[H]
	\begin{center}
		\includegraphics[width=.56\textwidth]{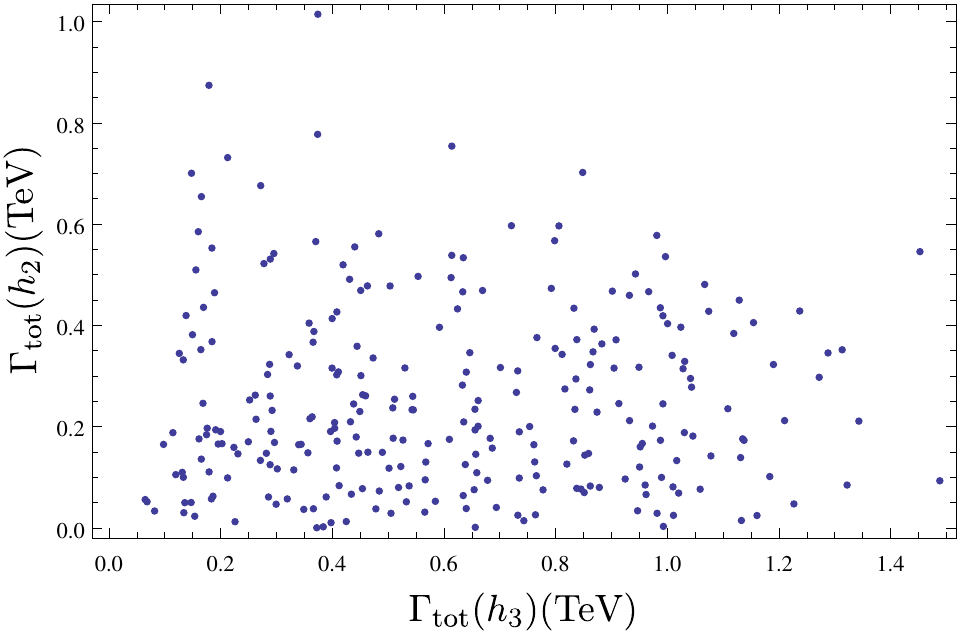}
		\caption{\label{fig:i9} The total decay width of the $h_{2}$ versus the one of $h_{3}$.}
	\end{center}
\end{figure}

\begin{table}[H]
	\tbl{\label{tab:12n} The branching ratios (BRs) of the heavy scalar $h_{2}$ for differnet channels.}
	{\begin{tabular}{@{}ccccccccccccccc@{}} \toprule
			Decay channel & BR where $m_{h_{2}}\in$ [500-1300](GeV) &  BR where $m_{h_{2}}\in$ [1300-3500](GeV)\\ \colrule
				$K_{1}^{+}K_{1}^{-}$&0& [$10^{-3}$-0.26]\\
			$U\bar{U}$&0&[$10^{-4}$-$10^{-2}$]\\
				$J\bar{J}$&0&[$10^{-4}$-$10^{-1}$]\\
			\botrule
	\end{tabular}}
\end{table}
	\newpage
\indent Now we turn our discussion to $h_{3}$ decay, Figure \ref{fig:i6} shows BR$(h_{3}\longrightarrow ZZ)$ as a function of the heavy Higgs $h_3$ mass for $v_\chi=v_\eta$=2 TeV where in the allowed parameter space, the mass of $m_{h_{3}}$ ranges from 3 TeV to 4 TeV. The decay channel $h_{3}\longrightarrow ZZ$ is the most important one with a BR $\geq 80\%$ which is about $\sim$0.8, while, the BR in $h_{1}h_{1}$ is the second most significant decay channel contributing roughly  $20\%$ as we shown in figure \ref{fig:h11}, whereas the lowest value of the branching ratio in that region is just below $\sim10^{-9}$.\\ \indent The branching ratios of the other processes are shown in the following table: 
\begin{table}[H]
	\tbl{\label{tab:1nm} The branching ratios (BRs) of the heavy scalar $h_{3}$ for differnet channels.}
	{\begin{tabular}{@{}ccccccccccccccc@{}} \toprule
			Decay channel & BR where $m_{h_{3}}\in$ [3000-4000](GeV)\\ \colrule
			$b\bar{b}$&$\sim \mathcal{O}(10^{-5})$\\
			$\tau^{+}\tau^{-}$&[$10^{-7}-10^{-9}$]\\
			$\gamma\gamma$&$[10^{-7}-10^{-5}]$\\
			$\gamma Z$&$[10^{-3}-10^{-2}]$\\
							$gg$&$[10^{-5}-10^{-4}$]\\
						ZZ& $\sim0.80$\\
			$Z^{\prime}Z^{\prime}$&$[10^{-7}-10^{-5}]$\\
			$K^{\prime0}K^{\prime0}$&[$1.5.10^{-6}-1.9.10^{-4}$]\\
			$Y^{+}Y^{-}$&[$6.10^{-6}-5.10^{-5}$]\\
			$X^{+}X^{-}$&[$10^{-6}-3.10^{-5}$]\\
			$V^{++}V^{--}$&[$10^{-6}-10^{-5}$]\\
			$K_{1}^{+}K_{1}^{-}$&[$1.5.10^{-6}-1.9.10^{-4}$]\\
			$U\bar{U}$& [$5.10^{-2}-0.25$]\\
			$J\bar{J}$&[$5.10^{-2}-10^{-3}$]\\
			$h_{1}h_{1}$&$\sim 0.20$\\
			$h_{1}h_{2}$&[$10^{-4}-10^{-2}$]\\
			$h_{2}h_{2}$&[$5.10^{-4}-10^{-3}$]\\
			\botrule
	\end{tabular}}
\end{table}
\indent At the end, we give the total decay width for the two CP-even scalars $h_{2}$ and $h_{3}$ in figure \ref{fig:i9}. It is well known that the SM Higgs with a mass of 125 GeV has a
very narrow width almost equal to $\Gamma_{h}\sim$ 4 MeV. \\
\indent In our case, the total width of $h_{2}$ is in the range 4.1.$10^{-5}$-1.014 (TeV), we notice a very narrow width of $h_{2}$ when the decay of $h_{2}$ into the heavy gauge bosons and exotic quarks are closed and it can decay only into a pair of leptons, quarks, scalars and $\gamma\gamma (Z\gamma)$, while, when they opened the total width is large and it achieves the value $\sim$ 1.014 TeV. Whereas, the total width of $h_{3}$ can be located between 0.0649 and 1.488 TeV. 
\section{Conclusion}
\label{sec:Conclusion}
~~~The compact 341 model is a gauge extension of the Standard Model with only three quadruplets scalar fields composed
by three neutral scalars, four singly charged and two doubly charged scalar bosons. In this work we discussed some phenomenology of the scalar sector. \\
\indent  We have used a set of the theoretical constraints such as: perturbative unitarity, boundedness from below constraints and the positivity of the scalar bosons masses condition to constrain the scalar parameters.\\
\indent  We have calculated the
signal strength of $h_1$ for different channels where we obtained a good fit to the data as can be seen in figure \ref{fig:i}. In
particular, with $\upsilon_{\chi}$= 2 TeV, we have shown that the compact 341 model is achieved to a good result that are compatible to the measurement of LHC.\\
\indent The second focus of our paper is the calculation of the branching ratio of the other heavy scalar bosons
$h_{2}$ and $h_{3}$, we found that $h_{2}$ decay
preferentially into a pair of Higgs-like particles with a branching ratio $\sim$ 1 when $m_{h_2} \in$ [500,1300] GeV, a feature not
easily obtained in other extensions of the Standard Model and in the second region where $m_{h_2} \in$ [1300,3500] GeV, $h_2$ prefers to decay into the heavy gauge bosons $X^{\mp},K_1^{\mp},K^{\prime0},V^{\mp\mp}$. Moreover, we found that $h_{3}$ decay
preferentially into a pair of Z bosons where $m_{h_3} \in$ [3000,4000] GeV .\\
\indent Finally, we have studied the total decay width of $h_{2}$ and $h_{3}$, we found that $\Gamma_{h_{2}}$ and $\Gamma_{h_{3}}$ can achieve 1.014 TeV and 1.488 TeV respectively. \\
\indent In conclusion, we have studied the scalar sector of the compact 341 model and showed that at a scale of a few TeV this model is a compelling alternative to the SM once, it is able to reproduce the experimental data
 regarding the signal strength.\\
\indent The production process $gg \longrightarrow h _{2,3}$ followed by
the decays $h_{2,3} \longrightarrow h_{1} h_{1}$ could be sizeable and could be an important source of double $h_{1}$ production where it is rather difficult to produce it using the conventional channel.
\appendix
\label{sec:Appendices}
\section{Perturbative unitarity and Vacuum stability conditions}
The perturbative unitarity conditions are \cite{Djouala:2020wjj}:
\begin{eqnarray}
\alpha^2\lambda_{4}&+&\gamma^2(\lambda_{6}+\lambda_{9})<16\pi,\nonumber\\
\beta^2\lambda_{4}&+&\sigma^2(\lambda_{6}+\lambda_{9})<16\pi,\nonumber\\
\alpha\beta\lambda_{4}&+&\gamma\sigma(\lambda_{6}+\lambda_{9})<8\pi,\nonumber\\
(\alpha^2&+&2\alpha\beta)\lambda_{4}+\gamma^2\lambda_{6}<32\pi,\nonumber\\
\lambda_{4}&+&\lambda_{6}<32\pi,\nonumber\\
\beta^2\lambda_{4}&+&\sigma^2\lambda_{6}<32\pi,\nonumber\\
\sigma\lambda_{9}&+&\beta\lambda_{7}<16\pi,\nonumber\\
\gamma\lambda_{9}&+&\alpha\lambda_{7}<16\pi,\nonumber\\
\gamma\sigma\lambda_{6}&<&16\pi,\nonumber\\
\alpha^2\lambda_{4}&+&\gamma^2\lambda_{6}+\alpha^2\lambda_{7}<16\pi,\nonumber\\
\beta^2\lambda_{4}&+&\sigma^2\lambda_{6}+\beta^2\lambda_{7}<16\pi,\nonumber\\
\alpha\beta\lambda_{4}&+&\sigma\gamma\lambda_{6}+\alpha\beta\lambda_{7}<8\pi,\nonumber\\
2\alpha^2\lambda_{1}&+&2\gamma^2\lambda_{3}+\lambda_{5}(\alpha^2+\gamma^2)+\lambda_{8}(\alpha^2+\gamma^2+(\sqrt{2}+1)\gamma\alpha)<32\pi,\nonumber\\
2\beta^2\lambda_{1}&+&2\sigma^2\lambda_{3}+\lambda_{5}(\beta^2+\sigma^2)+\lambda_{8}(\beta^2+\sigma^2+\sigma\beta)<32\pi,\nonumber\\
2\beta\alpha\lambda_{1}&+&4\sigma\gamma\lambda_{3}+2\lambda_{5}(\beta\alpha+\gamma^2)+\lambda_{8}(2\beta\alpha+2\sigma\alpha+\beta+\sigma\gamma+\beta\gamma+\sigma\beta)<32\pi\nonumber,\\
\lambda_{1}&+&\lambda_{3}+\lambda_{5}<32\pi,\nonumber
\end{eqnarray}
\begin{eqnarray}
\alpha^4\lambda_{1}&+&\gamma^2\lambda_{3}<32\pi,\nonumber\\
\beta^4\lambda_{1}&+&\sigma^4\lambda_{3}+\sigma^2\beta^2\lambda_{5}<32\pi,\nonumber\\
6\beta^2\alpha^2\lambda_{1}&+&6\sigma^2\gamma^2\lambda_{3}+\lambda_{5}(4\alpha\beta\gamma\sigma+\gamma^2\alpha^2+\gamma^2\beta^2+\alpha^2\sigma^2)<32\pi,\nonumber\\
2\alpha^3\beta\lambda_{1}&+&2\lambda_{3}\gamma^3\sigma+\lambda_{5}(\beta\alpha\gamma^2+\alpha^2\gamma\sigma)<16\pi,\nonumber\\
2\beta^3\alpha\lambda_{1}&+&2\lambda_{3}\sigma^3\gamma+\lambda_{5}(\beta\alpha\sigma^2+\beta^2\gamma\sigma)<16\pi,\nonumber\\
\lambda_{4}&+&\lambda_{6}+\lambda_{9}<16\pi,\nonumber\\
\lambda_{4}&+&\lambda_{6}+\lambda_{7}<16\pi.
\end{eqnarray}
The vacuum stability and the minimization conditions are satisfied only if \cite{Djouala:2020wjj}:
\begin{eqnarray}
\lambda_{1}&>&0,\qquad \lambda_{2}>0, \qquad\lambda_{3}>0,\nonumber\\ \lambda_{4}&+&2\sqrt{\lambda_{1}\lambda_{2}}>0\nonumber.\\ \lambda_{4}&+&\lambda_{7}+2\sqrt{\lambda_{1}\lambda_{2}}>0\nonumber\\
-2\sqrt{\lambda_{1}\lambda_{3}}&<&\lambda_{5}< 2\sqrt{\lambda_{1}\lambda_{3}},\nonumber\\
-2\sqrt{\lambda_{2}\lambda_{3}}&<&\lambda_{6}< 2\sqrt{\lambda_{2}\lambda_{3}},\nonumber\\
-2\sqrt{\lambda_{1}\lambda_{3}} &<&\lambda_{5}+\lambda_{8}< 2\sqrt{\lambda_{1}\lambda_{3}},\nonumber\\
-2\sqrt{\lambda_{2}\lambda_{3}}&<&\lambda_{6}+\lambda_{9}< 2\sqrt{\lambda_{2}\lambda_{3}}\nonumber,\\
4(\lambda_{4}&+&\lambda_{7})\lambda_{3}-2(\lambda_{5}+\lambda_{8})(\lambda_{6}+\lambda_{9})+2\sqrt{\Lambda_{1}}>0,\nonumber\\
4\lambda_{4}\lambda_{3}&-&2\lambda_{5}\lambda_{6}+2\sqrt{\Lambda_{2}}>0,\nonumber\\
4\lambda_{4}\lambda_{3}&-&2\lambda_{5}(\lambda_{6}+\lambda_{9})+2\sqrt{\Lambda_{3}}>0,\nonumber\\
4\lambda_{4}\lambda_{3}&-&2(\lambda_{5}+\lambda_{8})(\lambda_{6}+\lambda_{9})+2\sqrt{\Lambda_{4}}>0,\nonumber\\
4\lambda_{4}\lambda_{3}&-&2(\lambda_{5}+\lambda_{8})\lambda_{6}+2\sqrt{\Lambda_{5}}>0,\nonumber\\
4(\lambda_{4}&+&\lambda_{7})\lambda_{3}-2\lambda_{5}\lambda_{6}+2\sqrt{\Lambda_{6}}>0,\nonumber\\
4(\lambda_{4}&+&\lambda_{7})\lambda_{3}-2\lambda_{5}(\lambda_{6}+\lambda_{9})+2\sqrt{\Lambda_{7}}>0,\nonumber\\
4(\lambda_{4}&+&\lambda_{7})\lambda_{3}-2(\lambda_{5}+\lambda_{8})\lambda_{6}+2\sqrt{\Lambda_{8}}>0\label{eq:delenda2},
\end{eqnarray}
where:
\begin{eqnarray}
\Lambda_{1}&=&(4\lambda_{1}\lambda_{3}-(\lambda_{5}+\lambda_{8})^2)(4\lambda_{2}\lambda_{3}-(\lambda_{6}+\lambda_{9})^2),\nonumber\\
\Lambda_{2}&=&(4\lambda_{1}\lambda_{3}-\lambda_{5}^2)(4\lambda_{2}\lambda_{3}-\lambda_{6}^2),\nonumber\\
\Lambda_{3}&=&(4\lambda_{1}\lambda_{3}-\lambda_{5}^2)(4\lambda_{2}\lambda_{3}-(\lambda_{6}+\lambda_{9})^2),\nonumber\\
\Lambda_{4}&=&(4\lambda_{1}\lambda_{3}-(\lambda_{5}+\lambda_{8})^2)(4\lambda_{2}\lambda_{3}-(\lambda_{6}+\lambda_{9})^2),\nonumber\\
\Lambda_{5}&=&(4\lambda_{1}\lambda_{3}-(\lambda_{5}+\lambda_{8})^2)(4\lambda_{2}\lambda_{3}-\lambda_{6}^2),\nonumber\\
\Lambda_{6}&=&(4\lambda_{1}\lambda_{3}-\lambda_{5}^2)(4\lambda_{2}\lambda_{3}-\lambda_{6}^2),\nonumber\\
\Lambda_{7}&=&(4\lambda_{1}\lambda_{3}-\lambda_{5}^2)(4\lambda_{2}\lambda_{3}-(\lambda_{6}+\lambda_{9})^2),\nonumber\\
\Lambda_{8}&=&(4\lambda_{1}\lambda_{3}-(\lambda_{5}+\lambda_{8})^2)(4\lambda_{2}\lambda_{3}-\lambda_{6}^2).
\end{eqnarray}
\section*{Acknowledgments}
We are very grateful to the Algerian ministry of higher education and scientific research and DGRSDT for the financial support.

\end{document}